\documentclass[lettersize,journal]{IEEEtran}
\usepackage{tabularx}
\usepackage{multirow}
\usepackage{graphics} 
\usepackage{epsfig} 
\usepackage{mathptmx} 
\usepackage{times} 
\usepackage{amsmath} 
\usepackage{amssymb}  

\usepackage{amsmath,amsfonts}
\usepackage{algorithmic}
\usepackage{algorithm}
\usepackage{array}
\usepackage[caption=false,font=normalsize,labelfont=sf,textfont=sf]{subfig}
\usepackage{textcomp}
\usepackage{stfloats}
\usepackage{url}
\usepackage{verbatim}
\usepackage{graphicx}
\usepackage{cite}
\usepackage{float}
\usepackage{tikz}
\usetikzlibrary{shapes,arrows,calc, decorations.pathreplacing,calligraphy}
\definecolor{MyGreen}{RGB}{50,140,80}
\DeclareMathOperator{\arcsinh}{asinh}

\usepackage{accents}
\usepackage[english]{babel}
\makeatletter
\newcommand*{\rom}[1]{\expandafter\@slowromancap\romannumeral #1@}
\begin{document}
\newcommand{\Rl}[2]{\ensuremath{\mathbb{R}^{#1}_{#2}}}   
\newcommand{\R}{\ensuremath{\mathbb{R}}}
\newcommand{\Rlp}{\ensuremath{\mathbb{R}_{>0}}}
\newcommand{\Rlpc}{\ensuremath{\overline{\mathbb{R}}_{>0}}}
\newcommand{\Rlo}{\ensuremath{\mathbb{R}_{\geq 0}}}
\newcommand{\Rln}{\ensuremath{\mathbb{R}_{< 0}}}
\newcommand{\Zo}{\ensuremath{\mathbb{Z}_{\geq 0}}}
\newcommand{\Zp}{\ensuremath{\mathbb{Z}_{> 0}}}
\newcommand{\Np}{\ensuremath{\mathbb{N}_{> 0}}}
\newcommand{\N}{\ensuremath{\mathbb{N}}}                 
\newcommand{\Z}{\ensuremath{\mathbb{Z}}}

\definecolor{bleucit}{rgb}{0.1,0.4,0.8}
\newcommand{\bleucit}{\textcolor{bleucit}}
\newcommand{\postoyanbleucit}{\textcolor{bleucit}}

\newcommand\cp[1]{{`\emph{#1}'}}
\newcommand\blue[1]{\emph{{\color{blue}#1}}}

\newcommand{\cmark}{\ding{51}}%
\newcommand{\xmark}{\ding{55}}%

\newcommand{\dst}{\displaystyle}
\newcommand{\Linf}[1]{\ensuremath{\mathcal{L}^{#1}}}

\newcommand{\eg}{{\it e.g.}}

\newcommand{\Nesic}{Ne{\v{s}}i{\'c} }

\definecolor{blue_cv}{rgb}{0.09,0.35,0.78}

\newcommand{\KL}{\ensuremath{\mathcal{KL}}}
\newcommand{\K}{\ensuremath{\mathcal{K}}}
\newcommand{\Kinf}{\ensuremath{\mathcal{K}_{\infty}}}
\newcommand{\KK}{\ensuremath{\mathcal{KK}}}
\newcommand{\KN}{\ensuremath{\mathcal{KN}}}
\newcommand{\KKL}{\ensuremath{\mathcal{KKL}}}
\newcommand{\KLL}{\ensuremath{\mathcal{KLL}}}
\newcommand{\D}{\ensuremath{\mathcal{D}}}
\newcommand{\PD}{\ensuremath{\mathcal{PD}}}

\newcommand{\Cs}{\ensuremath{C_{\text{steady}}}}
\newcommand{\Ct}{\ensuremath{C_{\text{transient}}}}
\newcommand{\Ds}{\ensuremath{D_{\text{steady}}}}
\newcommand{\Dt}{\ensuremath{D_{\text{transient}}}}
\newcommand{\UtGpAS}{U$_{\text{t}}$GpAS}
\newcommand{\UtGAS}{U$_{\text{t}}$GAS}
\newcommand{\UjGpAS}{U$_{\text{j}}$GpAS}
\newcommand{\UjGAS}{U$_{\text{j}}$GAS}

\newcommand{\argmin}{\ensuremath{\text{argmin}\,}}
\newcommand{\interior}{\ensuremath{\text{int}\,}}
\newcommand{\dom}{\ensuremath{\text{dom}\,}}
\newcommand{\Span}{\ensuremath{\text{Span}}}
\newcommand{\avg}{\ensuremath{\text{avg}}}
\newcommand{\co}{\ensuremath{\text{co}\,}}
\newcommand{\coc}{\ensuremath{\overline{\text{dom}}\,}}
\newcommand{\ext}{\ensuremath{\text{ext}}}
\newcommand{\rge}{\ensuremath{\text{rge}\,}}
\newcommand{\esup}{\ensuremath{\text{ess.sup}\,}}

\newcommand{\sign}[1]{\ensuremath{\text{sign}{(#1)}}}
\newcommand{\sat}{\ensuremath{\text{sat}}}

\newcommand{\sinc}{\ensuremath{\text{sinc}}}
\newcommand{\nom}{\ensuremath{\text{nom}}}

\newcommand{\Tmati}{\ensuremath{T_{MATI}\,\,}}
\newcommand{\Tmasp}{\ensuremath{\mathrm{T_{MASP}}\,}}
\newcommand{\lc}{\ensuremath{\llbracket}}
\newcommand{\rc}{\ensuremath{\rrbracket}}

\newcommand{\norm}[1]{\ensuremath{\left\|{#1}\right\|}}
\newcommand{\ip}[2]{\ensuremath{\left\langle #1, #2\right\rangle}}
\newcommand{\cb}[1]{\ensuremath{\overline{\mathbb{B}}_{\mbox{\scriptsize $#1$}}}}                              
\newcommand{\ob}[2]{\ensuremath{\mathbb{B}_{\mbox{\scriptsize $#1$}}\ensuremath{\left( #2\right)}}}   
\newcommand{\df}{\ensuremath{\stackrel{\mbox{\tiny $\mathrm{def}$}}{=}\:}}                                             
\newcommand{\myint}[4]{\ensuremath{\int_{#1}^{#2}#3\;\mathrm{d}#4}}
\newcommand{\Mm}{\ensuremath{\:\stackrel{\rightarrow}{\scriptstyle{\rightarrow}}\:}}
\newcommand{\bm}[1]{\ensuremath{\mathbf{#1}}}
\newcommand{\hs}[1]{\hspace*{#1 em}}
\newcommand{\qa}{\ensuremath{\mathcal{Q}_{A}}}%
\newcommand{\mc}[1]{\ensuremath{\mathcal{#1}}}
\newcommand{\di}{\ensuremath{\mathcal{D}_{i}}}

\newcommand{\HS}{\ensuremath{\mathcal{H}}}
\newcommand{\HSc}{\ensuremath{\mathcal{H}_c}}


%

\newcommand{\ie}{{\it i.e. }}

\newtheorem{exple}{Example}
\newtheorem{defn}{Definition}
\newtheorem{claim}{Claim}
\newtheorem{hypo}{Hypothesis}
\newtheorem{ass}{Assumption}
\newtheorem{prop}{Proposition}
\newtheorem{fact}{Fact}
\newtheorem{lem}{Lemma}
\newtheorem{ex}{Example}
\newtheorem{thm}{Theorem}
\newtheorem{cor}{Corollary}
\newtheorem{pb}{Problem}
\newtheorem{rem}{Remark}
\newtheorem{sass}{Standing Assumption}


%
%
\newenvironment{rems}{\textit{Remarks. }}{\mbox{}\\[1ex]}

\title{Enhancing accuracy of finite-dimensional models
for lithium-ion batteries, observer design and
experimental validation}

\author{Mira Khalil$^{1,2}$, 
        Romain Postoyan$^{1}$ and
        St{\' e}phane Ra{\"e}l$^{2}$
\thanks{*This work was funded by Lorraine Universit{\' e} d’Excellence LUE.}
\thanks{$^{1}$Universit{\' e} de Lorraine, CNRS, CRAN, F-54000 Nancy, France.
{\tt(mira.khalil@univ-lorraine.fr,}
{\tt romain.postoyan@univ-lorraine.fr).}}
\thanks{$^{2}$Universit{\' e} de Lorraine, GREEN, F-54000 Nancy, France.
{\tt(stephane.rael@univ-lorraine.fr).}}
}




\maketitle

\begin{abstract}
Accurate estimation of the internal states of lithium-ion batteries is key towards improving their management for safety, efficiency and longevity purposes. Various approaches exist in the literature in this context, among which designing an observer based on an electrochemical model of the battery dynamics. With this approach, the performance of the observer depends on the accuracy of the considered model. It appears that electrochemical models, and thus their associated observer,  typically require to be of high dimension to generate accurate internal variables. In this work, we present a method to mitigate this limitation by correcting the lithium concentrations generated by a general class of finite-dimensional electrochemical models such that they asymptotically match those generated by the original partial differential equations (PDE) they are based on, for constant input currents. These corrections apply irrespectively of the order of the considered finite-dimensional model. The proposed correction leads to a new  state space model for which we design observers, whose global, robust convergences are supported by a Lyapunov analysis. Both numerical and experimental validations are presented, which show the improvement of the accuracy of the state estimates as a result of the proposed corrections.
\end{abstract}


\section{Introduction}

\IEEEPARstart{W}{ith} the increasing integration of applications that adopt lithium-ion batteries for their energy storage needs, ensuring their safe and efficient operation becomes of paramount importance. Proper monitoring of the battery state is thus needed, which can be achieved by the battery management system (BMS) provided it is fed with precise battery variables. Unfortunately, some key variables cannot be measured directly with sensors and they therefore need to be estimated. The state of charge (SOC), which is directly
related to the lithium concentrations in the battery electrodes, is one example of an unmeasurable key battery variable that needs to be estimated.

The battery state estimation problem has been thoroughly investigated in the literature e.g., \cite{HANNAN2017834,Barillas_2015, CHEMALI2018242, NG20091506, Klein2013TCST }. A common method is to design an observer based on a mathematical model of the battery internal dynamics e.g., \cite{Allam2018, AllamTCST2021, Blondel_IFAC_Journal_2017, Blondel_TCST_Journal_2019, 7004795, DreefDonkersCDC2018, Klein2013TCST, Moura2017TCST }. Several types of battery models are available for this purpose, see e.g., \cite{HE2012113,app8050659,WANG2020110015}. In particular, electrochemical models are suitable to describe the battery internal dynamics. These models are expressed by a set of partial differential equations (PDE), which describe the following phenomena: lithium diffusion within electrode active materials, electron migration in
 electrodes, ion migration and diffusion within the electrolyte and electrochemical kinetics of lithium insertion/de-insertion at electrode/electrolyte interface. More details about electrochemical models can be found in e.g., \cite{Doyle_1993,Fuller_1994}. Although these models can accurately generate
internal state variables, their mathematical structure is often too complex for observer design. For this reason, reduced electrochemical models are considered instead. One popular approach is assuming the particles within an electrode
behave like an average particle, we talk of single particle models (SPM) as in e.g., \cite{1998HaranBS, Blondel_TCST_Journal_2019,Domenico2010LithiumIonBS, 9920541,7004795,ScottMoura_ACC_2012}. Giving that solving PDEs analytically or numerically can be computationally demanding and complex, the PDEs are usually turned into ordinary differential
equations (ODE) via spatial discretization. A finite-dimensional model is thus obtained, which is convenient to design and implement an observer. Nevertheless, for the model to be faithful to the original PDEs, it typically needs to be of high dimension. This implies that the associated observer may also need to be of high dimension, which may make its design numerically challenging and may be an issue for its implementation.

In this work, we present a method to alleviate the need for finite-dimensional electrochemical models to be of high dimension to generate accurate variables. We consider for this purpose finite-dimensional SPMs, which include those in e.g., \cite{7004795, Blondel_TCST_Journal_2019,Allam2018,Domenico2010LithiumIonBS, 9920541,ScottMoura_ACC_2012} as special cases. We then present a technique to systematically correct the concentrations generated by these models so that these asymptotically match the concentrations given by the original PDEs for constant currents. Hence, for any given
model order, we obtain that the corrected concentrations
from the finite-dimensional models asymptotically tend to
the actual concentrations of the original infinite-dimensional
model for constant inputs thereby asymptotically eliminating
the errors induced by spatial discretization. Although the
purpose of these corrections is to eliminate asymptotic errors
for constant inputs, the provided simulation results show that
significant improvements may also be obtained for short time
horizons with a rapidly changing current profile. We then
exploit these corrected concentrations to derive a new output
voltage equation, which leads to a new state space model.

Afterwards, we present two methods to design an observer for the new, corrected model. The first method consists in assuming that an observer has already been designed for the original model without correction, using for instance the results of e.g., \cite{Blondel_IFAC_Journal_2017,Blondel_TCST_Journal_2019,7004795}, and we derive conditions under which the same observer structure still converges for the new model; we talk of observer emulation. If these conditions are not satisfied, an alternative is to directly design an observer for the new model. We present a method for this purpose, which is based on polytopic and Lyapunov-based tools similarly to e.g., \cite{Blondel_IFAC_Journal_2017,DreefDonkersCDC2018, zemouche2008}. This method guarantees the robust convergence of the state estimates generated by the observer to the actual battery states provided a linear matrix inequality holds. We then explain how to correct the estimated concentrations to asymptotically track those of the original PDEs in absence of disturbances and for constant inputs. Simulation results are presented to illustrate the improvements brought by the corrected model and the associated estimation schemes. An experimental validation of the obtained results is also provided, which shows that the cell voltage generated by the new, corrected model is improved by about 25$\%$ compared to the same model without correction, which results in an improved SOC estimate by about 25$\%$.

Compared to the preliminary version of this work \cite{KHALILCDC2023}, completely novel elements include: (i) the generalization of the considered class of SPM models, which captures more general spatial discretizations; (ii) the detailed analysis of the corrected concentrations; (iii) the emulation-based observer; (iv) the experimental validation of the results.

The rest of this paper is organised as follows. The considered class of SPM models is given in Section II. The correction of lithium concentrations is presented in Section $\text{\rom{3}}$. The new state space model is derived in Section $\text{\rom{4}}$. The observers designs are presented in Section $\text{\rom{5}}$. Numerical simulations are provided in
Section $\text{\rom{6}}$. The obtained results are validated experimentally in Section $\text{\rom{7}}$. Section $\text{\rom{8}}$ concludes the paper. All the parameters used in the paper are summarized
in Table \ref{param_table}.

\noindent \textbf{Notation.} Let $\mathbb{R}$ be the set of real numbers, $\mathbb{R}_{>0}:=(0,\infty)$, $\mathbb{R}_{\geq0}:=[0,\infty)$, $\mathbb{R}_{<0}:=(-\infty,0)$, $\mathbb{Z}$ be
the set of integers, $\mathbb{Z}_{>0}:=\{1,2,3,...\}$ and $\mathbb{C}$ be
the set of complex numbers. We use $\mathbb{I}_n$ to denote the identity matrix of dimension $n$, $\textbf{0}_{n \times m}$ the zero matrix of $\mathbb{R}^{n \times m}$ and $\textbf{1}_{n \times m}$ the matrix of $\mathbb{R}^{n \times m}$ whose elements are all equal to 1, with $n,m \in \mathbb{Z}_{>0} $. Given square matrices $A_1,..., A_n$, diag($A_1,..., A_n$) is the block diagonal matrix, whose block diagonal components are $A_1,..., A_n$ and $\underline{\text{diag}}(A_1,..., A_n)$ ($\overline{\text{diag}}(A_1,..., A_n)$) is the lower (upper) block diagonal matrix, whose lower (upper) block diagonal components are $A_1,..., A_n$. Given a real, symmetric matrix $P$, its maximum
and minimum eigenvalues are denoted by $\lambda_\text{max}(P)$ and
$\lambda_\text{min}(P)$ respectively. The symbol $*$ in a matrix stands for the symmetric term, i.e., $\left(\begin{smallmatrix}
    A & B \\ * & C 
\end{smallmatrix}\right)=\left(\begin{smallmatrix}
    A & B \\ B^\top & C 
\end{smallmatrix}\right)$. Given a vector $x \in \mathbb{R}^n$, $x^\top$ denotes the transpose of $x$. Given $x\in \mathbb{R}^{n}$ and $y\in \mathbb{R}^{m}$ with $n,m\in\mathbb{Z}_{>0}$, we use the notation $(x,y)$ to denote $(x^\top,y^\top)^\top$. Given two functions $f, g: \mathbb{C} \to \mathbb{C}$, we write $f(x)$ $\underset{x \to 0}{\sim}$ $g(x)$ when $ \lim_{x\to 0} \frac{f(x)}{g(x)}=1 $, in which case the functions $f$ and $g$ are said to be equivalent at 0. Given $f: \mathbb{R} \to \mathbb{R}^n$ with $n \in \mathbb{Z}_{>0}$, $(f)_\infty$ stands for $\lim_{t \to \infty}f(t)$ when it exists. For a vector $x \in \mathbb{R}^N$, $|x|$ denotes its Euclidean norm. For
a matrix $A \in \mathbb{R}^{n \times m}$, $\|A\|$ stands for its 2-induced norm,  ker($A$):=\{$x \in \mathbb{R}^m \,:\, Ax=\textbf{0}_{n}\}$ and for any $i \in \mathbb{R}^n$, $j \in \mathbb{R}^m$, $(A)_{ij}$ represents the $j$-th element of
the $i$-th row of matrix $A$. Let $f: \mathbb{R}_{\geq 0} \rightarrow \mathbb{R}^N, \|f\|_{\mathcal{L}_2,[0,t)}$ denotes the $\mathcal{L}_2$ norm of $f$ on the interval $[0,t)$, where $t \in [0, \infty)$. We write $f \in \mathcal{L}_2$, when $\|f\|_{\mathcal{L}_2,[0,\infty)} < \infty$.

\begin{table}[!h]
\caption{PARAMETER DEFINITION AND NUMERICAL VALUES }
\vspace{-0.6cm}
\label{parameters}
\begin{center}
\begin{tabular}{lll}
\hline
$A_\text{cell}$  & Cell area [$\text{m}^2$] & 0.8 \\
$F$ & Faraday constant [$\text{C.mol}^{-1}$] & 96485\\
$R$ & Gaz constant [$\text{J.K}^{-1}.\text{mol}^{-1}$] & 8.3145\\
$T$ & Temperature [K] & 298.15\\
$N_\text{pos}$ & Number of samples of the positive\\ =$N_\text{neg}$ & and negative electrode  & 4\\
$u_T$ & Thermal voltage ($u_T=\frac{RT}{F}$) [V] & \\
$d_\text{pos}$ & Thickness of the positive electrode [$\mu$m] & 36.4   \\
$d_\text{neg}$ & Thickness of the negative electrode [$\mu$m]&  50\\
$d_\text{sep}$ & Thickness of the separator [$\mu$m] &  25.4\\
$D_\text{pos}$ & Solid diffusion coefficient [m$^2$.s$^{-1}$] & $3.7 \times 10^{-16}$\\
$D_\text{neg}$ & Solid diffusion coefficient [m$^2$.s$^{-1}$] & $2 \times 10^{-16}$\\
$R_\text{pos}$ & Particle radius of positive electrode [$\mu$m] & 1  \\
$R_\text{neg}$ & Particle radius of negative electrode [$\mu$m] & 1 \\
$j_0^\text{pos}$ & Exchange current density of positive electrode \\ &[A.m$^{-2}$]& 0.54 \\
$j_0^\text{neg}$ & Exchange current density  of negative electrode\\ &[A.m$^{-2}$]& 0.75\\
$\epsilon_\text{pos}$ & Active material volume fraction [-]&  0.5\\
$\epsilon_\text{neg}$ & Active material volume fraction [-]&  0.58\\
$\epsilon_{e,\text{pos}}$ & Electrolyte phase volume fraction [-]&  0.33\\
$\epsilon_{e,\text{neg}}$ & Electrolyte phase volume fraction [-]&  0.332\\
$\epsilon_{e,\text{sep}}$ & Electrolyte phase volume fraction [-]&  0.5\\
$\sigma_\text{pos}$& Electronic conductivity [S/m]& 10 \\
$\sigma_\text{neg}$& Electronic conductivity [S/m]& 100\\
$\kappa_e$ & Ionic conductivity at 298.15 K [S/m] & 0.63\\
$Q_\text{cell}$ & Cell capacity [Ah] & 6\\
$Q$ & Lithium quantity in solid phase [Ah]& 11.396\\
$c_0^\text{pos}$ & Lithium concentration at $SOC = 0\%$ [$\text{mol}$.m$^{-3}$]& 25699\\
$c_0^\text{neg}$ & Lithium concentration at $SOC = 0\%$ [$\text{mol}$.m$^{-3}$]& 2199 \\
$c_{100}^\text{pos}$ & Lithium concentration at $SOC = 100\%$ [$\text{mol}$.m$^{-3}$]& 10324 \\
$c_{100}^\text{neg}$ &Lithium concentration at $SOC = 100 \%$ [$\text{mol}$.m$^{-3}$]& 11849 \\
$c_\text{max}^\text{pos}$ & Maximum concentration [$\text{mol}$.m$^{-3}$]& 29461 \\
$c_\text{max}^\text{neg}$ & Maximum concentration [$\text{mol}$.m$^{-3}$]& 17525\\

\hline
\end{tabular}
\end{center}
\label{param_table}
\end{table}

\section{Preliminaries on the class of spm models}

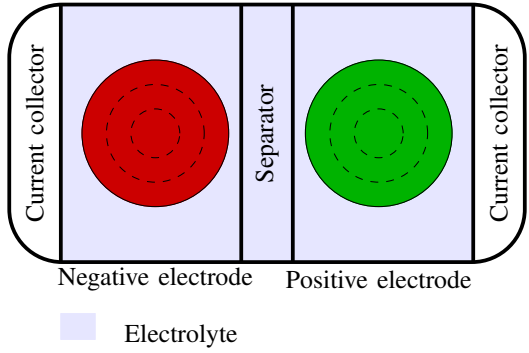
\begin{figure}
	\begin{center}

		\tikzstyle{blockBattery} = [draw,rounded corners=20,minimum height=15em, minimum width=30em,line width=0.5mm]

		\tikzstyle{blockElectrolyte} = [draw,minimum height=1em, fill=blue!7, minimum width=2em]
		
		\begin{tikzpicture}[auto, node distance=2cm,>=latex , scale=0.65,transform shape] 
			
			   \draw[draw=blue,color=white!90!blue,fill] (-12em,-7.5em) rectangle ++(24em,15em);
			\node [blockBattery] (battery) at (0,0) {};

             \draw[line width=0.5mm] (-1.5em,7.5em) -- (-1.5em,-7.5em);
             \draw[line width=0.5mm] (1.5em,7.5em) -- (1.5em,-7.5em);
             \node[scale=1.5,rotate=90] at (0,0) {Separator};
            \draw[line width=0.5mm] (-12em,7.5em) -- (-12em,-7.5em);
            \node[scale=1.5,rotate=90] at (-13.5em,0) {Current collector};
            \draw[line width=0.5mm] (12em,7.5em) -- (12em,-7.5em);
            \node[scale=1.5,rotate=90] at (13.5em,0) {Current collector};
            
            \draw[color=black!30!green,fill] (6.5em,0) circle [radius=1.5];
            \draw[color=black!20!red,fill] (-6.5em,0) circle [radius=1.5];

            \draw[color=black,line width=0.00001mm,dashed] (-6.5em,0) circle [radius=0.5];
            \draw[color=black,line width=0.00001mm,dashed] (-6.5em,0) circle [radius=1];
            \draw[color=black,line width=0.1mm] (-6.5em,0) circle [radius=1.5];
            
            \draw[color=black,line width=0.00001mm,dashed] (6.5em,0) circle [radius=0.5];
            \draw[color=black,line width=0.00001mm,dashed] (6.5em,0) circle [radius=1];
            \draw[color=black,line width=0.1mm] (6.5em,0) circle [radius=1.5];

			\node[scale=1.5] at (-6.5em,-8.5em) {Negative electrode};
            \node[scale=1.5] at (6.5em,-8.5em) {Positive electrode};

            \draw[draw=black,color=white!90!blue,fill] (-12em,-12em) rectangle ++(2em,1.5em);
             \node[scale=1.5] at (-5em,-11.8em) {Electrolyte};

		\end{tikzpicture}
	\end{center}
	\vspace*{-0.2cm} 
	\caption{\normalfont{Battery model schematic.}}
	\label{FIG:BatteryModel}
\end{figure}

We present in this section the class of SPM models, whose
lithium concentrations will then be corrected in Section III. We first briefly recall the main elements of a lithium-ion cell, namely: the positive electrode, the separator and the negative
electrode, which are all immersed in the electrolyte, and two current collectors, see Figure \ref{FIG:BatteryModel} for an illustration. The electrolyte is an ionic solution that ensures ion transport within the battery. The porous separator is an electrical insulator
that does not allow electrons to flow between the two electrodes. However, being porous, it allows the passage of ions via the electrolyte. The positive and negative electrodes consist of almost spherical particles of porous materials. The electrodes structure creates a surface of contact between the electrodes and the electrolyte producing electrochemical couples between them and thus introducing a potential difference between the positive and negative electrode. Further information about lithium-ion batteries can be found in e.g., \cite{Scrosati2004LithiumBS}.

We focus on  models ensuring the next assumption.

\begin{sass}[SA1]
The following holds: (i) lithium insertion or de-insertion reactions are homogeneous along the thickness of each electrode; (ii) the electrolyte dynamics is neglected; (iii) the temperature of the cell is constant and homogeneous.\hfill$\Box$
\label{SA1}
\end{sass}

Item (i) implies that each electrode can be reduced to a single particle, whose size is equal to the average size of all the particles that compose the actual electrode, we talk of SPM assumption, see e.g., \cite{ Domenico2010LithiumIonBS, ScottMoura_ACC_2012,7004795 ,Blondel_TCST_Journal_2019, 9920541,1998HaranBS}. As customarily done in electrochemical modeling of lithium-ion batteries, electrodes material particles are supposed spherical.
In item (ii), we ignore the electrolyte dynamics, which is reasonable for moderate currents and moderate temperatures. For high current rates and/or low temperatures, item (ii) can be relaxed and electrolyte dynamics can be added to the presented model and observers by applying the results of \cite{Bezine} \emph{mutatis mutandis}.
In item (iii), we suppose that the temperature is constant however, when the temperature varies, and is measured we can adapt the model and the developed observers of Section $\text{\rom{5}}$ to take into account the temperature variation like in \cite{Blondel_IFAC_Journal_2017}, \cite{9920541}. As for the temperature homogeneity assumption, it is reasonable at moderate and high temperatures. For low temperatures, it may become invalid and be interpreted as a parametric uncertainty, which can be handled by the observers of Section $\text{\rom{5}}$ if the uncertainty is small enough, like in \cite{9920541}.

Given SA1, the main dynamical phenomenon is the lithium diffusion in the electrodes active particles. This phenomenon is described using the next PDEs (see e.g., \cite{Doyle_1993}), for any $t\geq 0$ and $r \in [0,R_s]$, where $R_s>0$ is the radius of the particle in electrode $s \in \{\text{neg}, \text{pos}\}$, with neg and pos denoting the negative and positive electrode, respectively,
\begin{equation}
\begin{array}{lll}
        \varphi_s(r,t)&=&-D_s\frac{\partial c_s(r,t)}{\partial r}  \\
       \frac{\partial c_s(r,t)}{\partial t}&=& \frac{1}{r^2}\frac{\partial}{\partial r}\left(D_s r^2 \frac{\partial c_s(r,t)}{\partial r}\right),
\end{array}
\label{PDE}
\end{equation}
where $c_s$ is the local concentration of lithium, $\varphi_s$ is the lithium flux density and $D_s>0$ is the diffusion coefficient of lithium, along with the boundary conditions $\varphi_s (0,t)=0$ and $\varphi_s(R_s,t)=\frac{j^{\text{Li}}_s(t)}{a_sF}$, where  $j^{\text{Li}}_s \in \mathbb{R}$ is the electrochemical reaction
rate, $a_s:=\frac{3\epsilon_s}{R_s}$ is the active surface per volume unit, $\epsilon_s>0$ is the volume fraction of the active material particle and $F>0$ is Faraday’s constant.

To derive a set of ODEs from (\ref{PDE}), a spatial discretization method is performed. Hence, each particle is discretized into $N_s \in \mathbb{Z}_{>0}$ samples, where $s \in \{\text{neg},\text{pos}\}$. A zero-order approximation is made, i.e., we assume that the lithium concentration in each sample, $c_{s,n}$ for $n \in \{1, . . . , N_s\}$ and $s \in \{\text{neg},\text{pos}\}$, is homogeneous.
From the obtained set of ODEs, we derive the next state space equation, with the index $s \in \{\text{neg},\text{pos}\}$
\begin{equation}
\Dot{x}_s=A_sx_s+B_sm_s,
\label{SSODE}
\end{equation}
where $x_s:=(c_{s,1},\hdots,c_{s,{N_s}}) \in \mathbb{R}^{N_s}$ is the concatenation of the concentrations in electrode $s$ and $m_s:=-\frac{j^{\text{Li}}_s}{\epsilon_sF} \in \mathbb{R}$ is the input. The matrices $A_s \in \mathbb{R}^{N_s \times N_s} $ and $B_s \in \mathbb{R}^{N_s \times 1}$ are defined as $A_s:=\text{diag}(-\mu_1^s,-\upsilon_2^s,\hdots,-\upsilon_{N_s-1}^s,-\Tilde{\mu}_{N_s}^s)+\underline{\text{diag}}(\Tilde{\mu}_2^s,\hdots,\Tilde{\mu}_{N_s}^s)+\overline{\text{diag}}(\mu_1^s,\hdots,\mu_{N_s-1}^s)$, $B_s:=
    \begin{pmatrix}
    \textbf{0}_{1 \times (N_s-1) } & \frac{V_s}{V_{N_s}^s}
   \end{pmatrix}^\top $, where $\mu_{i}^s:=\frac{S_{i}^s}{r_{i+1}^s-r_i^s}\ \frac{D_s}{V_i^s}$
for any $i \in \{1,\hdots,N_s-1\}$, $\Tilde{\mu}_{i}^s:=\frac{S_{i-1}^s}{r_{i}^s-r_{i-1}^s}\frac{D_s}{V_i^s}$ for any $i \in \{2,\hdots,N_s\}$, $\upsilon_{i}^s:=\Tilde{\mu}_{i}^s+\mu_{i}^s$ for any $i \in \{2,\hdots,N_s-1\}$, $V_s:=\frac{4}{3}\pi (R_s)^3$ is the particle volume, $V_i^s:=\frac{4}{3}\pi (({r_i^s})^3-(r_{i-1}^s)^3)$ is the volume of sample $i$ and $S_i^s:=4\pi (r_i^s)^2$ its external surface with $r_i^s>0$ representing its external radius.

One of the lithium concentrations, namely $c_{\text{neg},1}$ is removed by exploiting the assumed lithium conservation, which is essential to design a converging observer. As a result, the dimension of the model is $N_\text{neg}+N_\text{pos}-1$. We will return to this point in Section $\text{\rom{4}.B}$.

\section{Concentrations correction}

The spatial discretization of (\ref{PDE}) to obtain (\ref{SSODE}) generates errors on the concentrations given by (\ref{SSODE}). These errors can be reduced by increasing the number of samples $N_s$, $s\in\{\text{neg},\text{pos}\}$, but this leads to a high-dimensional system in (\ref{SSODE}), which may lead to computational issues, especially when using the model in (\ref{SSODE}) for observer design. We present in this section an alternative method to reduce these errors by correcting the lithium concentrations generated by (\ref{SSODE}) so that they asymptotically match those given by (\ref{PDE}) for constant inputs irrespectively of the considered number of spatial samples, as formalized in Section $\text{\rom{3}.A} $. We can already emphasize that, although these corrections are established by considering the asymptotic behavior of (\ref{PDE}) and (\ref{SSODE}) for constant inputs, these may allow improving the accuracy of the concentrations given by (\ref{SSODE}) even for rapidly changing inputs as illustrated in Sections $\text{\rom{6}}$ and $\text{\rom{7}}$.
In this section, we denote by $c_{s,(1)}$ the lithium concentrations generated from the PDEs in (\ref{PDE}) and $c_{s,(2)}:=x_s$ the lithium concentrations generated by model (\ref{SSODE}) for electrode $s$, with $s\in\{\text{neg},\text{pos}\}$ for the sake of convenience.
\subsection{Main result}
We propose to correct the concentrations generated by model (\ref{SSODE}) as follows, for $j \in \{1,\hdots,N_s\}$ and $s\in\{\text{neg},\text{pos}\}$, 
\begin{equation}
    c_{s,\text{cor},j}:=c_{s,(2),\text{mean}}-K^{s}_j(c_{s,(2),\text{mean}}-c_{s,(2),j})
\label{conc_corr}
\end{equation}
where $c_{s,\text{cor}}$ represents the corrected lithium concentrations of model (\ref{SSODE}),  $c_{s,(2),\text{mean}}:=\frac{1}{V_s}\sum_{n=1}^{N_s}V^s_nc_{s,(2),n}$ is the lithium-ion mean concentration in electrode $s$ given by model (\ref{SSODE}) and $K^{s}_j \in \mathbb{R}$ is a static correction coefficient given by 
\begin{equation}
K^{s}_j:=
    \begin{cases}
        \frac{k(r^s_j)}{ ({\Tilde{A}_s}^{-1} \textbf{1}_{(N_s-1) \times 1})_{j1}}& j \in \{1,\hdots,N_s-1\}\\
       \frac{-k(r^s_j)V^s_{N_s}}{\Gamma_s^\text{red}{\Tilde{A}_s}^{-1} \textbf{1}_{(N_s-1) \times 1}}& j=N_s,
    \end{cases}
    \label{coeff_corr}
    \end{equation}
where $k(r_j^s):=\frac{\left(\frac{r_j^s}{R_s}\right)^2-\frac{3}{5}}{6}\tau_s$, $\tau_s:=\frac{R_s^2}{D_s}$, $\Gamma_s^\text{red}:=\begin{pmatrix}
    V^s_1 & V^s_2 &\hdots & V^s_{N_s-1}
\end{pmatrix} \in \mathbb{R}^{1 \times N_s-1}$, $r_j^s$ is defined at the end of Section II and $\Tilde{A}_s \in \mathbb{R}^{(N_s-1) \times (N_s-1)}$ is defined by $(\Tilde{A}_s)_{ij}:=(A_s)_{ij}-(A_s)_{iN_s}\frac{V^s_j}{V^s_{N_s}}$ for any $i,j \in \mathbb{R}^{(N_s-1)}$.

We next present the main result of this section.

\begin{thm}
For any constant input $m_s$ and any initial uniform profile for $c_{s,(1)}$, in the sense that there exists $c_0\in\R_{\geq 0}$ such that for any $r \in [0,R_s]$ $c_{s,(1)}(r,0)=c_0$, any corresponding solution $c_{s,(1)}$ to (\ref{PDE}) and $c_{s,(2)}$ to (\ref{SSODE}) with  $c_{s,(2)}(0)=\textbf{1}_{N_s \times 1}c_0$ satisfy for any $j \in \{1,\hdots,N_s\}$ and $s\in\{\text{neg},\text{pos}\}$
\begin{equation}
   ( c_{s,\text{cor},j}-c_{s,(1)}(r_j^s, \cdot))_\infty=0,
   \label{theorem1_result}
\end{equation} 
 where $c_{s,\text{cor}}$ is given in (\ref{conc_corr}) and $r_j^s$ is the external radius of each sample as defined after (\ref{SSODE}).
$\hfill \Box$
\label{T1}
\end{thm}
Theorem \ref{T1}
 implies that, as time tends to infinity, the corrected concentrations defined in (\ref{conc_corr}) match those generated by the PDEs in (\ref{PDE}) when the input is constant and $c_{s,(1)}(\cdot,0)$ is uniform. It is important to note that Theorem \ref{T1} imposes no conditions on the number of samples with which  the PDEs in (\ref{PDE}) are discretized, and thus no conditions on the dimension of (\ref{SSODE}) for (\ref{theorem1_result}) to hold. 

 To prove Theorem \ref{T1}, we first analyze properties of the lithium concentrations of (\ref{PDE}) in Section III.B and of the mean concentrations of (\ref{SSODE}) in Section III.C. We then investigate the error between the concentrations generated by (\ref{PDE}) and (\ref{SSODE}) in ``steady state'' in Section III.D. Lastly, we combine these properties to prove Theorem \ref{T1} in Section III.E. 

 \begin{rem}
      In Theorem \ref{T1}, the lithium concentrations $c_{s,(1)}$ have an initial uniform profile. We claim that this is not restrictive as there are always periods of time in the life of a lithium-ion battery where $c_{s,(1)}$ is homogeneous in electrode $s$, with $s \in \{\text{neg},\text{pos}\}$. $\hfill \Box$ 
 \end{rem}

\begin{figure}
     \centering
     \includegraphics[scale=0.27]{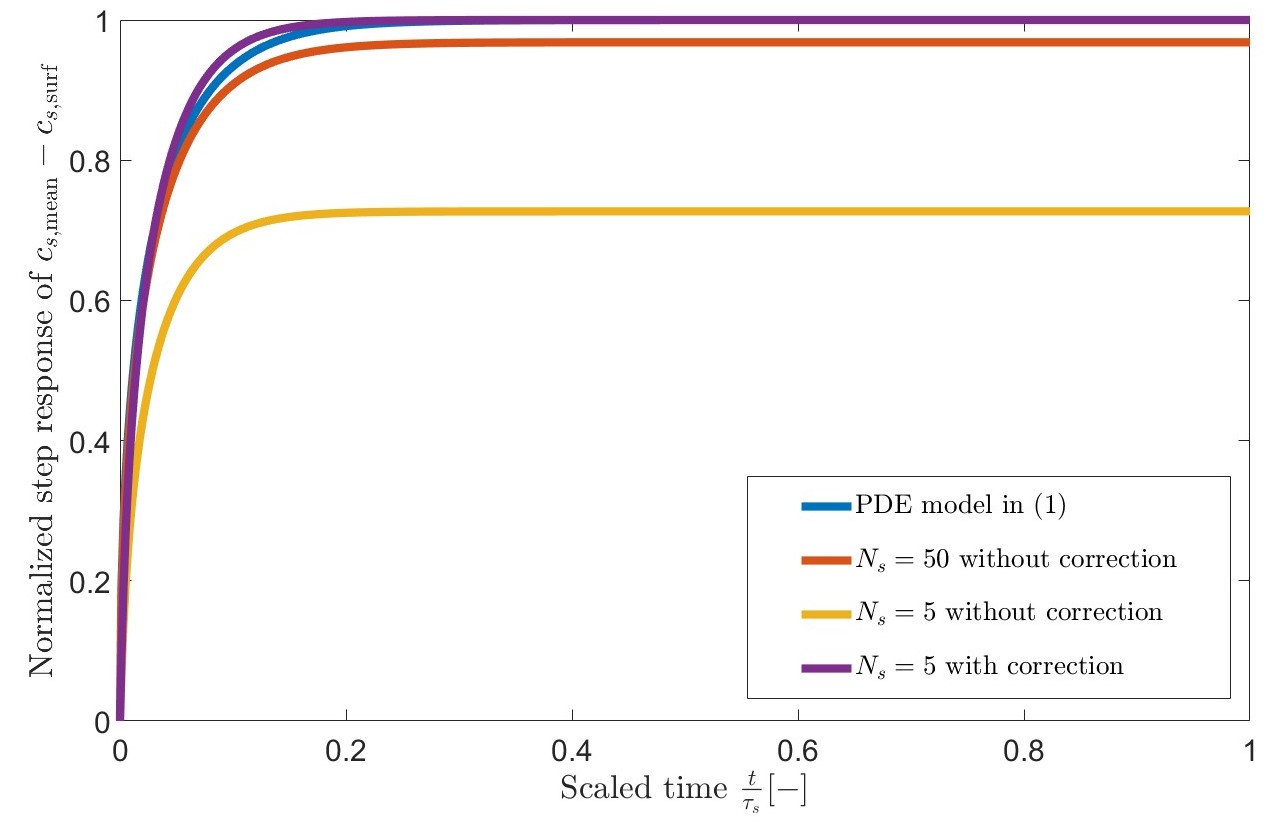}
     \caption{Normalized step response of $c_{s,\text{mean}}-c_{s,\text{surf}}$, $s \in \{\text{neg}, \text{pos}\}$ and $c_{s,\text{surf}}$ represents the surface concentration of electrode $s$, for the PDE model in (\ref{PDE}) and for model (\ref{SSODE}) with and without concentrations corrections.}
     \label{response}
 \end{figure}

 \begin{rem}
The surface concentrations correction is illustrated in Figure \ref{response}, where the normalized step response of $c_{s,\text{mean}}-c_{s,\text{surf}}$,
with $c_{s,\text{surf}}$ denoting the surface concentrations, is depicted for the PDE model in (\ref{PDE}), taken as the reference model,
and for model (\ref{SSODE}) with and without concentrations correction for different values of $N_s$. The error between the reference model
and model (\ref{SSODE}) without correction is significant for $N_s = 5$.
This error is reduced when increasing $N_s$ to 50. On the other hand, this error is drastically reduced and eliminated in steady state for
model (\ref{SSODE}) with correction by only taking  $N_s = 5$ samples. Further numerical illustrations of the advantages of the proposed corrections are provided in Sections $\text{\rom{6}}$ and $\text{\rom{7}}$. $\hfill \Box$ 
 \end{rem}

\subsection{Properties of the concentrations for model (\ref{PDE})}
The mean concentration of the PDEs in (\ref{PDE}), denoted as ${c}_{s,(1),\text{mean}}$, is obtained by integrating (\ref{PDE}) over the particle volume and using the corresponding boundary conditions. Thus, we derive that for any $t \geq 0$
\begin{equation}
   {\dot {c}_{s,(1),\text{mean}}(t)} = m_s(t),
   \label{conc_mean_dyn}
\end{equation}
with $m_s(t):=-\frac{j^{\text{Li}}_s(t)}{\epsilon_sF}$ as after (\ref{SSODE}).

We consider a uniform initial profile of lithium concentration in solid phase as in Theorem \ref{T1}, i.e., there exists $c_0\in\R_{\geq 0}$ such that for any $r \in [0,R_s]$ ${c}_{s,(1)}(r,0)=c_0$, with $c_0$ being the initial mean concentration ${c}_{s,(1),\text{mean}}(0)$. Let $c_{s,(1)}$ be a corresponding solution to (\ref{PDE}), we derive in the Laplace domain (see \cite{kremer}) that, for any $r \in [0,R_s]$ and $p \in \mathbb{C}$ the Laplace variable, \begin{equation}
\begin{array}{lll}
   \mathcal{C}_s(r,p)&=&F_s(r,p)m_s(p)  \\
    F_s(r,p)&:=&\frac{\tau_s}{3}\frac{\sinh{\left(r\sqrt{\frac{p}{D_s}}\right)}\frac{R_s}{r}}{\sqrt{\tau_s p}\cosh{(\sqrt{\tau_s p})}-\sinh{(\sqrt{\tau_s p}})},
\end{array}
\label{solutionpde}
\end{equation}
where $\mathcal{C}_s(r,p)$ is the Laplace transform of $c_{s,(1)}(r,t)-c_{s,(1)}(r,0)$, $m_s(p)$ the Laplace transform of $m_s(t)$ and $\tau_s:=\frac{R_s^2}{D_s}$ is the diffusion time constant.
It follows from \cite{kremer} that for any $r \in [0,R_s]$
\begin{equation}
     F_{s}(r,p)  \underset{p \rightarrow{}0}{\sim} \frac{1}{p}+k(r),
     \label{response_eq}
\end{equation}
where $k(r):=\frac{(\frac{r}{R_s})^2-\frac{3}{5}}{6}\tau_s$ is a constant as seen after (\ref{coeff_corr}).

In view of (\ref{solutionpde}) and (\ref{response_eq}), we obtain, for a constant input $m_s$, $\mathcal{C}_s(r,p)\underset{p \rightarrow{}0}{\sim}\frac{m_s(p)}{p}+k(r)m_s(p)$. On the other hand, the Laplace transform of (\ref{conc_mean_dyn}) gives $\mathcal{C}_{s,\text{mean}}(p)=\frac{m_s(p)}{p}$, with $\mathcal{C}_{s,\text{mean}}$ the Laplace transform of $c_{s,(1),\text{mean}}(t)-c_{s,(1),\text{mean}}(0)$. Hence, $\mathcal{C}_s(r,p)\underset{p\rightarrow{}0}{\sim} \mathcal{C}_{s,\text{mean}}(p)+k(r)m_s(p)$. In the time domain, for a constant input $m_s$ and $c_{s,(1)}(r,0)={c}_{s,(1),\text{mean}}(0)$, we derive that for any $r \in [0,R_s]$
\begin{equation}
    (c_{s,(1)}(r,\cdot))_\infty=({c}_{s,(1),\text{mean}})_\infty+k(r)m_s.
\label{prop_conc1}
\end{equation}

In view of (\ref{prop_conc1}), we deduce that, for any $r \in [0,R_s]$, the step response of the lithium concentrations of (\ref{PDE}) tends to the mean concentration of model (\ref{PDE}) up to an additional constant $k(r)m_s$. This property is exploited in Section III.D to evaluate the errors between the concentrations generated by (\ref{PDE}) and (\ref{SSODE}). Before that, a property of the mean concentration for model (\ref{SSODE}) is established in the next section.

\subsection{Properties of the mean concentration for model (\ref{SSODE}) }
The mean lithium concentration in electrode $s$ of model (\ref{SSODE}) is defined as
\begin{equation}
    c_{s,(2),\text{mean}}:=\frac{1}{V_s}\sum_{n=1}^{N_s}V^s_ic_{s,(2),n}=\frac{1}{V_s}\Gamma_s x_s,
\label{conc2_mean}
\end{equation}
with $\Gamma_s:=\begin{pmatrix}
    V^s_1 & V^s_2 &\hdots & V^s_{N_s}
\end{pmatrix} \in \mathbb{R}^{1 \times N_s}$. We have the next result on $c_{s,(2),\text{mean}}$.

\begin{lem}
For any Lebesgue measurable, locally essentially bounded input $m_s$, any corresponding solution $x_s$ to (\ref{SSODE}) satisfies $\Dot{c}_{s,(2),\text{mean}}(t)=m_s(t)$ for all $t \geq 0$.
$\hfill \Box$
\label{L1}
\end{lem}
\textbf{Proof.} Let $m_s$ be a Lebesgue measurable, locally essentially bounded input and $x_s$ be a corresponding solution to (\ref{SSODE}). In view of (\ref{SSODE}) and (\ref{conc2_mean}), we have $\Dot{c}_{s,(2),\text{mean}}=\frac{1}{V_s}\Gamma_s \dot x_s=\frac{1}{V_s}\Gamma_s(A_sx_s+B_sm_s)$. On the other hand, matrix $A_s$ in (\ref{SSODE}) satisfies $\Gamma_sA_s=\textbf{0}_{1 \times N_s}$. In addition, we have $\Gamma_sB_s=V_s$. Therefore, we obtain $\Dot{c}_{s,(2),\text{mean}}(t)=m_s(t)$ for all $t \geq 0$. $\hfill \blacksquare$

Lemma \ref{L1} implies that the mean lithium concentration given by model (\ref{SSODE}) is equal to the mean concentration of the PDEs in (\ref{PDE}) when these two are initialized at the same value i.e., when $c_{s,(1),\text{mean}}(0)=c_{s,(2),\text{mean}}(0)$, $c_{s,(1),\text{mean}}(t)=c_{s,(2),\text{mean}}(t)$ for any $t \geq 0$. The SPM model in \ref{SSODE} is thus conservative with respect to mean concentrations, see \cite{kremer}. We therefore use the short notation $c_{s,\text{mean}}$ in the following.

A consequence of this conservation property is that, by evaluating the normalized step response of $c_{s,\text{mean}}-c_{s}$ generated by (\ref{PDE}) and (\ref{SSODE}), respectively, we obtain in steady state, provided it exists as we show next, a constant error. We present in the next section a method to determine this error, which we use to correct the concentrations as in (\ref{conc_corr}).

\subsection{Steady state errors}
We define $\Tilde{x}_s:=x_{s,\text{mean}}-x_s$ the mismatch between $x_s$ and the vector of mean concentration $x_{s,\text{mean}}:={c}_{s,\text{mean}} \textbf{1}_{N_s \times 1}$. Using (\ref{SSODE}), Lemma \ref{L1} and the fact that $A_s\textbf{1}_{N_s \times 1}={\textbf{0}}_{N_s \times 1}$, the dynamics of $\Tilde{x}_s$ is given by
\begin{equation}
    \Dot{\Tilde{x}}_s=A_s\Tilde{x}_s+(\textbf{1}_{N_s \times 1}-B_s)m_s.
    \label{mismatch_ss}
\end{equation}

The next lemma allows to write any element $\tilde x_{s,i}$ of any solution $\tilde x_s=(\tilde x_{s,1 },\ldots,\tilde x_{s,{N_s}})$ to (\ref{mismatch_ss}) as a linear combination of all the others elements.
\begin{lem}
For any Lebesgue measurable, locally essentially bounded input $m_s$, any corresponding solution $\Tilde{x}_s$ to (\ref{mismatch_ss}) satisfies $\sum_{i=1}^{N_s} {V^s_i\Tilde{x}_{s,i}}(t)=0$ for all $t \geq 0$.$\hfill \Box$
\label{L2}
\end{lem}

\textbf{Proof.} Let $m_s$ be a Lebesgue measurable, locally essentially bounded input and $\Tilde{x}_s$ be a corresponding solution to (\ref{mismatch_ss}). From (\ref{mismatch_ss}), we obtain $\sum_{i=1}^{N_s}V^s_i\dot{\Tilde{x}}_{s,i}=\Gamma_s\dot{\Tilde{x}}_s=\Gamma_sA_s\Tilde{x}_s+\Gamma_s(\textbf{1}_{N_s \times 1}-B_s)m_s$. Given that $\Gamma_sA_s=\textbf{0}_{1 \times N_s}$ and $\Gamma_sB_s=\Gamma_s \textbf{1}_{N_s \times 1}$ as we have seen in the proof of Lemma \ref{L1}, we obtain $\sum_{i=1}^{N_s}V^s_i\dot{\Tilde{x}}_{s,i}(t)=0$ for all $t \geq 0$. Consequently, $\sum_{i=1}^{N_s}V^s_i{\Tilde{x}}_{s,i}(t)=\sum_{i=1}^{N_s}V^s_i{\Tilde{x}}_{s,i}(0)=0$ for all $t \geq 0$.
$\hfill \blacksquare$

Using Lemma \ref{L2}, we derive from (\ref{mismatch_ss}),
\begin{equation}
    \Dot{\Tilde{x}}_{s,\text{red}}=\Tilde{A}_s\Tilde{x}_{s,\text{red}}+\textbf{1}_{(N_s-1) \times 1}m_s,
    \label{reduced_ss}
\end{equation}
where $\Tilde{x}_{s,\text{red}}$ represents the first $N_s-1$ components of $\Tilde{x}_{s}$.
Matrix $\Tilde{A}_s \in \mathbb{R}^{N_s-1 \times N_s-1}$ is defined as in Section III.A after (\ref{coeff_corr}) and satisfies the next result.

\begin{lem}
$\Tilde{A}_s$ is Hurwitz.$\hfill \Box$
\label{L3}
\end{lem}
The proof of Lemma \ref{L3} is provided in Appendix to avoid breaking the flow of exposition.

Lemma \ref{L3} implies that, for a constant input $m_s$, any solution $\Tilde{x}_{s,\text{red}}$ to (\ref{reduced_ss}) converges to a constant value, thereby proving the existence of steady states. For any given constant input $m_s$, the steady state of $\Tilde{x}_{s,\text{red}}$ satisfies $(\Dot{\Tilde{x}}_{s,\text{red}})_\infty=(\Tilde{A}_s\Tilde{x}_{s,\text{red}}+\textbf{1}_{(N_s-1) \times 1}m_s)_\infty=\textbf{0}_{(N_s-1) \times 1}$. Therefore, we obtain
\begin{equation}
   (\Tilde{x}_{s,\text{red}})_\infty=-\tilde A_s^{-1}\textbf{1}_{(N_s-1) \times 1}m_s,
   \label{13}
\end{equation}
noting that $\Tilde{A}_s$ is invertible being Hurwitz by Lemma \ref{L3}.
In view of (\ref{13}), $(c_{s,\text{mean}}-c_{s,(2),j})_\infty=-({\Tilde{A}_s}^{-1} \textbf{1}_{(N_s-1) \times 1})_{j1}m_s$ for any $j \in \{1,\hdots,N_s-1\}$. On the other hand, we have from (\ref{prop_conc1}) that $(c_{s,\text{mean}}-c_{s,(1)}(r,\cdot))_\infty=-k(r)m_s$ for all $r \in [0,R_s]$. Hence, we obtain that for any $j \in \{1,\hdots,N_s-1\}$
\begin{equation}
    (c_{s,(2),j}-c_{s,(1)}(r_j^s,\cdot))_\infty=-k(r_j^s)m_s+({\Tilde{A}_s}^{-1} \textbf{1}_{(N_s-1) \times 1})_{j1}m_s.
    \label{14}
\end{equation}
It remains to calculate the step response in steady state of $c_{s,(2),{N_s}}-c_{s,(1)}(r_{N_s}^s,\cdot)$. In view of Lemma \ref{L2}, $\tilde x_{s,{N_s}}=-\frac{1}{V^s_{N_s}}\Gamma_s^\text{red}\tilde x_{s,\text{red}}$, $\Gamma_s^\text{red}$ represent the $N_s-1$ first elements of $\Gamma_s$ and defined after (\ref{coeff_corr}). Given (\ref{13}), we derive 
\begin{equation}
  (\Tilde{x}_{s,{N_s}})_\infty=\frac{1}{V^s_{N_s}}\Gamma_s^\text{red}\Tilde{A}_s^{-1}\textbf{1}_{(N_s-1) \times 1}m_s.
  \label{15}
\end{equation}
On the one hand, we have from (\ref{15}) that $(c_{s,\text{mean}}-c_{s,(2),{N_s}})_\infty=\frac{1}{V^s_{N_s}}\Gamma_s^\text{red}\Tilde{A}_s^{-1}\textbf{1}_{(N_s-1) \times 1}m_s$. On the other hand, we have from (\ref{prop_conc1}) that $(c_{s,\text{mean}}-c_{s,(1)}(r,\cdot))_\infty=-k(r)m_s$ for all $r \in [0,R_s]$. Therefore, we obtain 
\begin{multline}
     (c_{s,(2),{N_s}}-c_{s,(1)}(r_{N_s}^s,\cdot))_\infty=-k(r_{N_s}^s)m_s\\
     -\frac{1}{V^s_{N_s}}\Gamma_s^\text{red}\Tilde{A}_s^{-1}\textbf{1}_{(N_s-1) \times 1}m_s.
\label{16}
\end{multline}

To match the concentrations of model (\ref{SSODE}) to those of the PDEs in (\ref{PDE}) in steady state, the corrected concentrations of model (\ref{SSODE}) need to satisfy $(c_{s,\text{cor},j}-c_{s,(1)}(r_j^s,\cdot))_\infty=0$ for all $j \in \{1,\hdots,N_s\}$.
For this purpose, we introduce the correction coefficient $K^{s}_j$ defined in (\ref{coeff_corr}) such as $(c_{s,\text{mean}}-c_{s,\text{cor},j})_\infty=K^{s}_j(c_{s,\text{mean}}-c_{s,(2),j})_\infty$.
In the next section, we prove that the corrected concentrations match those of (\ref{PDE}) as stated in Theorem \ref{T1}.

\subsection{Proof of Theorem \ref{T1}}
We are now ready to prove Theorem \ref{T1}. Let $m_s$ be a constant input, $c_0\in\R_{\geq 0}$ and $c_{s,(1)}$ the solution to (\ref{PDE}) with $c_{s,(1)}(r,0)=c_0$ for any $r \in [0,R_s]$. As a consequence, $c_{s,\text{mean}}(0)=c_0$. Let $c_{s,(2)}$ be the corresponding solution for model (\ref{SSODE}) with input $m_s$ and initial condition $c_0 \textbf{1}_{N_s \times 1}$ .
In view of (\ref{conc_corr}) and the property of the lithium concentrations of (\ref{PDE}) presented in (\ref{prop_conc1}), we have, for any $j \in \{1,\hdots,N_s\}$,
\begin{multline}
( c_{s,\text{cor},j}-c_{s,(1)}(r_j^s,\cdot))_\infty\\
        =\left(c_{s,(2),\text{mean}}-K^{s}_j(c_{s,(2),\text{mean}}-c_{s,(2),j})-{c}_{s,(1),\text{mean}}\right.\\
        \left.-k(r_j^s)m_s \right)_\infty.
        \label{17}
\end{multline}
Given the conservation property of the mean concentrations of model (\ref{SSODE}) with respect to model (\ref{PDE}) established in Section III.C, (\ref{17}) is equivalent to, for any $j \in \{1,\hdots,N_s\}$, 
\begin{multline}
( c_{s,\text{cor},j}-c_{s,(1)}(r_j^s,\cdot))_\infty\\
        =(-K^{s}_j(c_{s,\text{mean}}-c_{s,(2),j})-k(r_j^s)m_s)_\infty.
\end{multline}
Using the steady state error calculated in (\ref{13}), we obtain for
$j \in \{1,\hdots,N_s-1\}$
\begin{multline}
( c_{s,\text{cor},j}-c_{s,(1)}(r_j^s,\cdot))_\infty\\
        =K^{s}_j\tilde A_s^{-1}\textbf{1}_{(N_s-1) \times 1}m_s-k(r_j^s)m_s.
        \label{19}
\end{multline}
As for $j=N_s$, in view of (\ref{15}), we derive
\begin{multline}
( c_{s,\text{cor},j}-c_{s,(1)}(r_j^s,\cdot))_\infty\\
        =-K^{s}_j\textbf{1}_{1 \times (N_s-1)}\tilde A_s^{-1}\textbf{1}_{(N_s-1) \times 1}m_s-k(r_j^s)m_s.
        \label{20}
\end{multline}
Given the expression of $K_j^s$ in (\ref{coeff_corr}), we obtain from (\ref{19}) and (\ref{20}), for any $j \in \{1,\hdots,N_s\}$, $( c_{s,\text{cor},j}-c_{s,(1)}(r_j^s,\cdot))_\infty=0$ as in (\ref{theorem1_result}). 
This completes the proof.


\section{State space model}

In this section, we exploit the results obtained in Section III to establish a new output equation for model (\ref{SSODE}), and thus a new state space model. Before that, the  relation between $m_s$ and the cell current $I_\text{cell}$ is recalled. Then, a model reduction is performed as it is essential to ensure the system detectability, which is (implicitly) exploited later in Section V like in \cite{Blondel_TCST_Journal_2019,7004795, 9920541,Klein2013TCST}.
\subsection{Relation between $m_s$ and $I_\text{cell}$}
Given SA\ref{SA1}, the electrochemical reaction rate is homogeneous within each electrode. Therefore, a proportional relationship can be established between $I_\text{cell}$ and $j^\text{Li}_s$ in particular $j^\text{Li}_\text{neg}:=\frac{I_\text{cell}}{A_\text{cell}d_\text{neg}}$ and $j^\text{Li}_\text{pos}:=-\frac{I_\text{cell}}{A_\text{cell}d_\text{pos}}$, where $I_ \text{cell}$ is in generator convention (i.e., $I_\text{cell}>0$ in discharge), $A_\text{cell}$ is the electrode surface and $d_s$ is the thickness of electrode $s$.
On the other hand, we have $m_s:=-\frac{j^\text{Li}_s}{\epsilon_sF}$. Hence, we obtain $
       m_\text{neg}:=   -\frac{I_\text{cell}}{\epsilon_\text{neg}A_\text{cell}d_\text{neg}F}$ and $
        m_\text{pos}:=  \frac{I_\text{cell}}{\epsilon_\text{pos}A_\text{cell}d_\text{pos}F}.$

\subsection{Model reduction}
Model (\ref{SSODE}) is reduced just like in e.g., \cite{Blondel_TCST_Journal_2019, 9920541, Klein2013TCST,7004795} by adopting the next assumption, which is
essential later for the observer convergence.
\begin{sass}[SA2]
    The quantity of lithium inserted in battery electrodes is constant and known. $\hfill \Box$
    \label{SA2}
\end{sass}
SA\ref{SA2} is reasonable over short periods of time. Factors such as cell degradation or side reactions can cause capacity loss over time, resulting in a reduction in the total quantity of lithium and the violation of SA\ref{SA2}. In this case, if there is a small uncertainty regarding the quantity of lithium, the battery state and its estimation would exhibit asymptotic small errors, as shown by the Lyapunov-based proof in Theorems \ref{T2} and \ref{T3} in Section $\text{\rom{5}}$. Conversely, if the uncertainty is big and thus the quantity of lithium needs to be estimated, state of health estimation algorithms may be employed as in e.g., \cite{AllamTCST2021,GAO2022}.

SA\ref{SA2} allows to write 
a lithium mass conservation. Hence, the quantity of lithium is defined as
\begin{equation}
    Q:=\alpha_\text{neg}\sum_{i=1}^{N_\text{neg}}c_{\text{neg},i}V^\text{neg}_i+\alpha_\text{pos}\sum_{i=1}^{N_\text{pos}}c_{\text{pos},i}V^\text{pos}_i,
    \label{lithium_consv}
\end{equation}
where $\alpha_s:=\frac{F}{3600}\frac{\epsilon_sA_\text{cell}d_s}{V_s}$ and $V_\text{s}$ is the volume of the particle of electrode $s$.
From (\ref{lithium_consv}), we express the lithium concentration at the center of the negative electrode
$c_{\text{neg},1}$ as a linear combination of all the other sampled concentrations in solid phase
\begin{equation}
    c_{\text{neg},1}=\overline{K}-\frac{1}{V^\text{neg}_1}\sum_{i=2}^{N_\text{neg}}c_{\text{neg},i}V^\text{neg}_i-\frac{\alpha_\text{pos}}{\alpha_\text{neg}V^\text{neg}_1}\sum_{i=1}^{N_\text{pos}}c_{\text{pos},i}V^\text{pos}_i,
    \label{model-red}
\end{equation}
where $\overline{K}:=\frac{Q}{\alpha_\text{neg}V^\text{neg}_1}$.

In view of (\ref{model-red}), $c_{\text{neg},1}$ is no longer needed in the state space representations as it can be recovered from the other concentrations.

\subsection{Corrected output equation}

We are ready to present the new output voltage equation.
The output equation of model (\ref{SSODE}) is obtained by decomposition of the cell voltage $V_\text{cell}$. The main components of $V_\text{cell}$ are the potential differences between the electrodes and the electrolyte called open circuit voltages (OCV) denoted $OCV_s$ for $s \in \{\text{neg}, \text{pos}\}$, which depend on the surface insertion rates $\zeta_s$ defined by $\zeta_s:=\frac{c_{s,{N_s}}}{c_\text{max}^s}$ for $s \in \{\text{neg},\text{pos}\}$, where $c_\text{max}^s$ is the maximum lithium concentration of electrode $s$ and $c_{s,N_s}$ is the surface concentration generated by model (\ref{SSODE}).
Given the correction of the lithium concentrations made in Section III, instead of using $c_{s,N_s}$ to define the surface insertion rates, we use the corrected surface concentration $c_{s,\text{cor},N_s}$ defined in (\ref{conc_corr}) to derive the corrected surface insertion rates $\zeta_{s,\text{cor}}$. As a result, we obtain the output equation for $y:=V_\text{cell}$
\begin{equation}
    y=OCV_\text{pos}(\zeta_{\text{pos},\text{cor}})-OCV_\text{neg}(\zeta_{\text{neg},\text{cor}})+g(u), 
    \label{new_out_eq}
\end{equation}
where $\zeta_{s,\text{cor}}:=\frac{c_{s,\text{cor},N_s}}{c_\text{max}^s}$ and $g(u):=-\eta_{r,\text{pos}}(u)-\eta_{\text{pos}}(u)-\eta_{\text{neg}}(u)-\eta_{r,\text{neg}}(u)-\eta_{r,\text{sep}}(u)$ for any $u:=I_\text{cell} \in \mathbb{R}$, where $\eta_s(u):=2\frac{RT}{F}\arcsinh \left( {\frac{R_s}{6j_0^s\epsilon_sA_\text{cell}d_s}u}\right)$, $\eta_{r,s}(u):=\frac{1}{2A_\text{cell}}\left(\frac{d_s}{\sigma_{s,\text{eff}}}+ \frac{d_s}{\kappa_s}\right)u$, $\eta_{r,\text{sep}}(u):=\frac{1}{A_\text{cell}}\frac{d_\text{sep}}{\kappa_\text{sep}}u$, $\kappa_s:=\kappa_e\epsilon_{e,s}^{1.5}$, $\sigma_{s,\text{eff}}:=\sigma_s\epsilon_s$, $\kappa_\text{sep}:=\kappa_e\epsilon_{e,\text{sep}}^{1.5}$, with $R$, $T$, $j_0^s$, $\kappa_e$, $\epsilon_{e,s}$, $\epsilon_{e,\text{sep}}$, $\epsilon_s$ and $\sigma_s$ defined in Table \ref{param_table}.

\begin{rem}
The output equation in e.g., \cite{Allam2018,Blondel_TCST_Journal_2019,7004795} is given by 
$y=OCV_\text{pos}(\zeta_{\text{pos}})-OCV_\text{neg}(\zeta_{\text{neg}})+g(u)$.
The difference between this output equation and the one in (\ref{new_out_eq}) is in the determination of the OCVs. In (\ref{new_out_eq}), the OCVs are determined from the corrected surface concentrations as in the other output equation they are determined directly from the surface concentrations generated by (\ref{SSODE}). The term $g(u)$, on the other hand, remains the same for both output equations.$\hfill \Box$
\end{rem}
\subsection{State space form}
We present the overall state space representation. We introduce for this purpose the state vector $x:=(c_{\text{neg},2},...,c_{\text{neg},N_\text{neg}},c_{\text{pos},1},...,c_{\text{pos},N_\text{pos}}) \in \mathbb{R}^N$ with $N:=N_\text{neg}-1+N_\text{pos}$ (recall that $c_{\text{neg},1}$ has been removed in Section \text{\rom{4}}.B), the input $u=I_\text{cell} \in \mathbb{R}$, the output $y=V_{\text{cell}} \in \mathbb{R}$, and $w \in \mathbb{R}^{n_w}$ and $v \in \mathbb{R}^{n_v}$ represent additive exogenous
perturbations and measurement noise respectively. We derive the next state space equation
\begin{equation}
    \begin{cases}
       \Dot{x}=Ax+Bu+K+Ew \\
    y=h_\text{cor}(x)+g(u)+v,
    \end{cases}
    \label{new_overallss}
\end{equation}
where $A:=\begin{pmatrix}
    A_1 & A_c \\
    A_r & \text{diag}(A_\text{neg}^\text{red},A_\text{pos})
\end{pmatrix}\in \mathbb{R}^{N \times N}$, 
$A_{1}:=\begin{pmatrix}-\upsilon_2^\text{neg}-\frac{V^\text{neg}_2}{V^\text{neg}_1}\mu_1^\text{neg}\end{pmatrix} \in \mathbb{R}^{1 \times 1}$,
$A_c:=$ $\left(\begin{smallmatrix}
\mu_2^\text{neg}-\frac{V^\text{neg}_3}{V^\text{neg}_1}\mu_1^\text{neg}&   &  -\frac{V^\text{neg}_{N_\text{neg}}}{V^\text{neg}_1}\mu_1^\text{neg}  &\hdots & -\frac{\alpha_\text{pos}V^\text{pos}_1}{\alpha_\text{neg}V^\text{neg}_1}\mu_1^\text{neg}& \hdots
&  -\frac{\alpha_\text{pos}V^\text{pos}_{N_\text{pos}}}{\alpha_\text{neg}V^\text{neg}_1}\mu_1^\text{neg}
\end{smallmatrix}\right)$ $\in \mathbb{R}^{1 \times (N-1)}$,
$A_r:=\begin{pmatrix}
\Tilde{\mu}_3^\text{neg}& \textbf{0}_{1 \times (N-2)}
\end{pmatrix}^\top \in \mathbb{R}^{(N-1) \times 1}$,
$A_\text{neg}^\text{red}:=\text{diag}(-\upsilon_3^\text{neg},\hdots,-\upsilon_{N_\text{neg}-1}^\text{neg},-\Tilde{\mu}_{N_\text{neg}}^\text{neg})+\underline{\text{diag}}(\Tilde{\mu}_4^\text{neg},\hdots,\Tilde{\mu}_{N_\text{neg}}^\text{neg})+\overline{\text{diag}}(\mu_3^\text{neg},\hdots,\mu_{N_\text{neg}-1}^\text{neg}) \in \mathbb{R}^{(N_\text{neg}-2) \times (N_\text{neg}-2)}$, $ B:=
    \begin{pmatrix}
    \textbf{0}_{1 \times (N_\text{neg}-2)} & -\overline{K}_I^\text{neg} & \textbf{0}_{1 \times (N_\text{pos}-1)} & \overline{K}_I^\text{pos}
    \end{pmatrix}^\top \in \mathbb{R}^{N \times 1}$,  $\overline{K}_I^s:=\frac{{V_s}}{V^s_{N_s}\epsilon_s FA_\text{cell}d_s}$ and $K:=
    \begin{pmatrix}
      \mu_{1}^\text{neg}  \overline{K}  & \textbf{0}_{1 \times (N-1)}
    \end{pmatrix}^\top \in \mathbb{R}^{N \times 1}$. The function $h_\text{cor}: \mathbb{R}^N \rightarrow \mathbb{R}$ is defined as, for any $x \in \mathbb{R}^N$, $h_\text{cor}(x) := OCV_\text{pos}(H_{\text{pos},\text{cor}}x)-OCV_\text{neg}(H_{\text{neg},\text{cor}}x+K_1)$, such that $\zeta_{\text{pos},\text{cor}}:=H_{\text{pos},\text{cor}}x$ and $\zeta_{\text{neg},\text{cor}}:=H_{\text{neg},\text{cor}}x+K_1$,
with $H_{s,\text{cor}} \in \mathbb{R}^{1 \times N}$ defined as
\begin{equation}
    \begin{array}{l}
         H_{\text{pos},\text{cor}}:=\begin{pmatrix}
           \textbf{0}_{1 \times (N_\text{neg}-1)} & \overline{h}^\text{pos}_1 & \hdots  & \overline{h}^\text{pos}_{N_\text{pos}-1} & \overline{h}^\text{pos}_{N_\text{pos}} +\frac{K^{\text{pos}}_{N_\text{pos}}}{c_\text{max}^\text{pos}}
           \end{pmatrix} \\
         H_{\text{neg},\text{cor}}:=\begin{pmatrix}
            \textbf{0}_{1 \times (N_\text{neg}-2)} & \frac{K^{\text{neg}}_{N_\text{neg}}}{c_\text{max}^\text{neg}} & -{h}_1^* & \hdots &-{h}_{N_\text{pos}}^*
           \end{pmatrix},
    \end{array}
\end{equation}
 where $\overline{h}^s_i:= \frac{V^s_i}{V_s c_\text{max}^s}(1-K^{s}_{N_s})
    $, ${h}_i^*:= \frac{V^\text{pos}_i}{V_\text{neg} c_\text{max}^\text{neg}}(1-K^{\text{neg}}_{N_\text{neg}})\frac{\alpha_\text{pos}}{\alpha_\text{neg}}$ and the constant $K_1:=\overline{h}^\text{neg}_1\overline{K}$.
    
We are now ready to proceed with the observer design for system (\ref{new_overallss}).

\section{State estimation}

In this section, we propose two methods to design a state observer for system (\ref{new_overallss}), which both rely on an assumption made on the OCVs presented in Section $\text{\rom{5}.A}$. The first one consists in assuming that an observer has been designed for the original model without correction and to derive conditions under which the same observer structure will be guaranteed to converge for the corrected model in (\ref{new_overallss}), see Section $\text{\rom{5}.B}$. If these conditions appear not to be satisfied, an alternative method is to directly design an observer for system (\ref{new_overallss}). We propose a polytopic based approach for this purpose in Section $\text{\rom{5}.C}$, similarly to e.g., \cite{Blondel_IFAC_Journal_2017, zemouche2008, DreefDonkersCDC2018}. We then explain how to correct the obtained state estimates, given by any of the two observers, to asymptotically match the concentrations given by the original PDEs in (\ref{PDE}) for constant inputs by exploiting the results of Section $\text{\rom{3}}$.
\subsection{Assumption on the OCVs}
We make the next assumption on the OCVs as in e.g., \cite{Blondel_IFAC_Journal_2017,DreefDonkersCDC2018,9920541}.
\begin{ass}
    For any $s \in \{\text{neg},\text{pos}\}$, there exist constant matrices $C_{s,1}, C_{s,2}  \in \mathbb{R}$ such as for any $z$, $z'$ $\in \mathbb{R}$,
    \begin{equation}
    OCV_s(z)-OCV_s({z'})=C_s(z,z')(z-z'),
    \label{ocv_ass1}
\end{equation}
where $C_s(z,z'):=\lambda^s_1(z,z')C_{s,1}+\lambda^s_2(z,z')C_{s,2}$ with $\lambda^s_i(z,z') \in [0,1]$ for $i \in \{1,2\}$ and $\lambda^s_1(z,z')+\lambda^s_2(z,z')=1$.$\hfill \Box$ 
\label{A1}
\end{ass}
Assumption \ref{A1} means that each $OCV_s$ lies in a polytope defined by $C_{s,1}, C_{s,2}$ with $s \in \{\text{neg},\text{pos}\}$. This condition is often
verified in practice. Indeed, the OCVs are
generally defined on the interval $[0, 1]$
and are typically well-approximated by a piecewise continuously
differentiable and thus globally Lipschitz function. Then, it suffices to extrapolate the OCVs on $[1,\infty)$ (resp. on $(-\infty,0$]) by using zero order or first order approximations based on the value of the OCVs at 1 (resp. at 0) for Assumption \ref{A1} to hold. Then, $C_{s,1}$ and  $C_{s,2}$ represent the minimum and maximum slopes of $OCV_s$, respectively. This is the case for the OCVs considered in Sections VI and VII, see Figure \ref{OCV}.

\subsection{Emulated observer}
In this section, we derive conditions under which an observer designed for the original model without correction can still be applied for the corrected model (\ref{new_overallss}). We first consider for this purpose the model without correction, given by
    \begin{equation}
    \begin{cases}
       \Dot{x}=Ax+Bu+K+Ew \\
    y=h(x)+g(u)+v,
    \end{cases}
    \label{oldss}
\end{equation}
where the function $h: \mathbb{R}^N \rightarrow \mathbb{R}$ is defined, for all $x \in \mathbb{R}^N$ by $h(x):=OCV_\text{pos}(H_\text{pos}x)-OCV_\text{neg}(H_\text{neg}x)$ as in Remark 3, with $H_\text{pos}:=\begin{pmatrix}
    \textbf{0}_{1 \times N-1} & \frac{1}{c_\text{max}^\text{pos}}
\end{pmatrix}$ and $H_\text{neg}:=\begin{pmatrix}
    \textbf{0}_{1 \times N_\text{neg}-2} & \frac{1}{c_\text{max}^\text{neg}} & \textbf{0}_{1 \times N_\text{pos}}
\end{pmatrix}$.

The designed observer is of the form
\begin{equation}
    \begin{cases}
        \Dot{\Hat{x}}=A\Hat{x}+Bu+K+L(y-\hat{y}) \\
        \hat{y}=h(\hat{x})+g(u),
        \label{PierreObs}
    \end{cases}
\end{equation}
where $\hat{x} \in \mathbb{R}^N$ is the state vector estimate, $L \in \mathbb{R}^N$ is the observation matrix gain and $\hat{y}$ is the estimated output. We assume that observer (\ref{PierreObs}) is designed to satisfy the next Lyapunov properties.
\begin{ass}
There exist $P$, $Q\in \mathbb{R}^{N \times N}$ symmetric, positive definite matrices and $\mu_w$, $\mu_v \in \mathbb{R}_{>0}$ such that for any $x, \hat x \in \mathbb{R}^N$, $w \in \mathbb{R}^{n_w}$ and $v \in \mathbb{R}^{n_v}$, denoting $V(e_1):=e_1^\top Pe_1$ and $e_1:=x-\hat x$, 
\begin{multline} 
   \left\langle\mathcal{r} V(e_1), Ae_1+Ew-L(h(x)-h(\hat{x}))-Lv\right\rangle\\ \leq -e_1^\top Qe_1 + \mu_w |w|^2 + \mu_v |v|^2.
   \label{eq_ass2}
   \end{multline}
$\hfill \Box$
\label{A2}
\end{ass}
Assumption \ref{A2} implies that system (\ref{oldss}), (\ref{PierreObs}) is $\mathcal{L}_2$-stable from $(w,v)$ to $e_1:=x-\hat x$, in particular that there exist $c \geq 0$ and $\varepsilon \in \mathbb{R}_{> 0}$ such that for any $w, v \in \mathcal{L}_2 $ and any $u$ Lebesgue measurable and locally essentially bounded, any  solution $(x,\hat x)$ to (\ref{oldss}), (\ref{PierreObs}) satisfies $\|e_1\|_{\mathcal{L}_2,[0,t)} \leq c|e_1(0)|+\sqrt{\frac{\mu_w}{\epsilon}} \|w\|_{\mathcal{L}_2,[0,t)}+\sqrt{\frac{\mu_v}{\epsilon}} \|v\|_{\mathcal{L}_2,[0,t)} $ for any $t \geq 0$. Moreover, when $w=0$ and $v=0$, $\{(x,\hat x) \,:\, x=\hat x\}$ is uniformly globally exponentially stable, i.e., there exist $\gamma_1\geq 1$, $\gamma_2 \in \mathbb{R}_{> 0}$ such that for any $u$ Lebesgue measurable and locally essentially bounded, any solution $(x,\hat x)$ to (\ref{oldss}), (\ref{PierreObs}) satisfies $|e_1(t)| \leq \gamma_1 |e_1(0)| e^{-\gamma_2 t}$ for any $t \geq 0$.
Examples of observer designs ensuring the satisfaction of Assumption \ref{A2} for the non-corrected model (\ref{oldss}) include, e.g., \cite{Blondel_IFAC_Journal_2017,Blondel_TCST_Journal_2019,7004795}.

When feeding observer (\ref{PierreObs}) with the new output equation (\ref{new_out_eq}), its convergence is no longer guaranteed in general. Our goal in this section is to identify sufficient conditions under which the same observer converges for model (\ref{new_overallss}). 
We thus consider the next observer, which is observer (\ref{PierreObs}) fed with the output equation (\ref{new_out_eq})  
\begin{equation}
    \begin{cases}
       \Dot{\hat{x}}=A\hat{x}+Bu+K+L(y-\hat{y}) \\
    \hat{y}=h_\text{cor}(\hat{x})+g(u),
    \end{cases}
    \label{newobs}
    \end{equation}
Note that the difference with (\ref{PierreObs}) is that $\hat y$ is defined using $h_\text{cor}$ defined after (\ref{new_overallss}) instead of $h$.
We define the estimation error $e:=x-\hat{x}$, whose dynamics  follows from the difference of the dynamics of (\ref{new_overallss}) and (\ref{newobs}) as follows
\begin{equation}
    \Dot{e}=Ae+Ew-L(h_\text{cor}(x)-h_\text{cor}(\hat{x}))-Lv.
    \label{error}
\end{equation}
By adding and subtracting the term $L(h(x)-h(\hat{x}))$ to (\ref{error}), we obtain
\begin{equation}
    \Dot{e}=Ae+Ew-L(h(x)-h(\hat{x}))-Lv+L(\Tilde{h}(x)-\Tilde{h}( \hat{x})),
    \label{error2}
\end{equation}
where $\Tilde{h}:=h-h_\text{cor}$.

A consequence of Assumption \ref{A1} is that the term $\Tilde{h}(x)-\Tilde{h}(\hat x)$ appearing in (\ref{error2}) can written as, for any $x, x' \in \mathbb{R}^N$
\begin{equation}
    \Tilde{h}(x)-\Tilde{h}(x')=\Tilde{C}(x,x')(x-x'),
    \label{33}
\end{equation}
where $\Tilde{C}(x,x'):=\sum_{i=1}^4 \lambda_i(x,x')\Tilde{C}_i$, with $\lambda_i(x,x') \in [0,1]$ for $i \in \{1, 2, 3, 4\}$ and $\sum_{i=1}^4 \lambda_i(x,x')=1$.
This means that $\Tilde{h}$ lies in a polytope defined by the vertices $\Tilde{C}_i$ with $i \in \{1, 2, 3, 4\}$, which are given in (\ref{v1}).

The next theorem presents the conditions under which observer (\ref{newobs}) converges for system (\ref{new_overallss}).

\begin{thm}
Suppose the following holds.
\begin{enumerate}
    \item[(i)] Assumptions \ref{A1} and \ref{A2} are satisfied.
    \item[(ii)] Matrices $P$ and $Q$ in Assumption \ref{A2} satisfy
    \begin{equation}
    -Q+\tilde C_i^\top L^\top P +PL \tilde C_i < 0,
    \label{c2_th2}
\end{equation}
where $\tilde C_i$, with $i\in \{1,2,3,4\}$, defined in (\ref{v1}).
\end{enumerate}
Then,
\begin{itemize}
    \item system (\ref{new_overallss}), (\ref{newobs}) is $\mathcal{L}_2$-stable from ($w,v$) to $e$ with gain less or equal to $\sqrt{\frac{\mu_w}{\epsilon}}$ and $\sqrt{\frac{\mu_v}{\epsilon}}$, respectively, where $\mu_w, \mu_v \in \mathbb{R}_{>0}$ come from Assumption \ref{A2} and $\varepsilon>0$ is any constant satisfying $-Q+\tilde C_i^\top L^\top P +PL \tilde C_i \leq -\varepsilon I_N$, in particular, there
exists $c \geq 0$ such that for any $w, v \in \mathcal{L}_2 $ and $u$ Lebesgue measurable and locally essentially bounded input, any  solution $(x,e)$ to (\ref{new_overallss}), (\ref{error}) satisfies $\|e\|_{\mathcal{L}_2,[0,t)} \leq c|e(0)|+\sqrt{\frac{\mu_w}{\epsilon}} \|w\|_{\mathcal{L}_2,[0,t)}+\sqrt{\frac{\mu_v}{\epsilon}} \|v\|_{\mathcal{L}_2,[0,t)} $ for any $t \geq 0$.

\item  $\{(x,e): e=0\}$ is uniformly globally exponentially stable when $w$=0 and $v$=0, i.e. there exist $\gamma_1\geq 1$, $\gamma_2 \in \mathbb{R}_{> 0}$ such that for any $u$ Lebesgue measurable and locally essentially bounded input, any solution $(x,e)$ to (\ref{new_overallss}), (\ref{error}) satisfies $|e(t)| \leq \gamma_1 |e(0)| e^{-\gamma_2 t}$ for any $t \geq 0$.$\hfill \Box$
\end{itemize}
\label{T2}
\end{thm}

\textbf{Proof.} 
Let $x, \hat{x} \in \mathbb{R}^{N}$, $w \in \mathbb{R}^{n_w}$ and $v \in \mathbb{R}^{n_v}$.
We consider $V(e)=e^\top Pe$, where $e:=x-\hat{x} \in \mathbb{R}^N$ as in Assumption \ref{A2}. We have $\lambda_\text{min}(P)|e|^2\leq V(e)\leq\lambda_\text{max}(P)|e|^2$ with $0<\lambda_\text{min}(P)\leq \lambda_\text{max}(P)$ as $P$ is symmetric, positive definite by Assumption \ref{A2}. In view of (\ref{error2}),
\begin{multline} 
   \left\langle\mathcal{r} V(e), Ae+Ew-L(h(x)-h(\hat{x}))-Lv+L(\Tilde{h}(x)-\Tilde{h}( \hat{x}))\right\rangle\\
  =2(Ae+Ew-L(h(x)-h(\hat x))-Lv)^\top Pe\\+2(L(\tilde h (x)- \tilde h (\hat x)))^\top Pe,
  \label{35}
\end{multline}
where we recall that $\tilde h=h-h_\text{cor}$. In view of Assumption \ref{A2}, 
\begin{multline} 
   2(Ae+Ew-L(h(x)-h(\hat x))-Lv)^\top Pe  \\
   \leq -e^\top Qe + \mu_w |w|^2 + \mu_v |v|^2.
   \label{36}
   \end{multline}

By substituting the term $\tilde h (x)- \tilde h (\hat x)$ in (\ref{35}) by its expression in (\ref{33}) and by using (\ref{36}), we derive, omitting the argument of $\lambda_i$, 
\begin{multline}
    \langle \mathcal{r} V(e), Ae+Ew-L(h(x)-h(\hat{x}))-Lv+L(\Tilde{h}(x)-\Tilde{h}( \hat{x})) \rangle\\
    \leq \sum_{i=1}^4 \lambda_i\left(-e^\top Qe + \mu_w |w|^2 + \mu_v |v|^2+(L \tilde C_i e)^\top Pe \right. \\ \left.+
   e^\top P L \tilde C_i e\right),
\end{multline}
recall that $\sum_{i=1}^4 \lambda_i=1$. Given that matrices $P$ and $Q$ in Assumption \ref{A2} satisfy (\ref{c2_th2}), there exist $\varepsilon \in \mathbb{R}_{>0}$ such that $-Q+\tilde C_i^\top L^\top P +PL \tilde C_i \leq -\varepsilon I_N$. We thus obtain
\begin{multline}
\langle \mathcal{r} V(e), Ae+Ew-L(h(x)-h(\hat{x}))-Lv+L(\Tilde{h}(x)-\Tilde{h}( \hat{x})) \rangle\\
   \leq e^\top(-Q+\tilde C_i^\top L^\top P +PL \tilde C_i)e + \mu_w |w|^2 + \mu_v |v|^2 \\
    \leq -\varepsilon |e|^2+ \mu_w |w|^2 + \mu_v |v|^2.
    \label{38}
\end{multline}
For any Lebesgue measurable and locally essentially bounded input current, the solutions to (\ref{new_overallss}) are defined for all positive time as the right hand side of (\ref{new_overallss}) is affine. We also have that for any Lebesgue measurable and locally essentially bounded input current $u$ and input $y$, and any $w, v \in \mathcal{L}_2$, system (\ref{newobs}) is forward complete as the only nonlinearity appearing in the right hand-side of (\ref{newobs}) is due to $h_\text{cor}$, which is globally Lipschitz as a consequence of Assumption \ref{A1}; see \cite[Theorem 3.2]{khalil2002nonlinear}.
In view of (\ref{38}), for any $w, v \in \mathcal{L}_2$ and any solution $x$ to (\ref{new_overallss}) and $\hat x$ to (\ref{newobs}), $e=x-\hat x$ verifies for all $t\geq 0$
\begin{equation*}
\dot V(e(t))\leq-\varepsilon|e(t)|^2+\mu_w |w(t)|^2+\mu_v |v(t)|^2,
\end{equation*}
which gives by \cite[Lemma 3.4]{khalil2002nonlinear} 
\begin{multline*}
 V(e(t)) \leq V(e(0))-  \varepsilon \int_0^t  |e(\tau)|^2d\tau+ \mu_w \int_0^t |w(\tau)|^2d\tau \\
+\mu_v \int_0^t |v(\tau)|^2d\tau.\\
\end{multline*}
Consequently, as $V(e(t)) \geq 0$, $\sqrt{\int_0^t |e(\tau)|^2d\tau } \leq \sqrt{\frac{V(e(0))}{\varepsilon} + \frac{\mu_w}{\varepsilon} \int_0^t |w(\tau)|^2d\tau  + \frac{\mu_v}{\varepsilon} \int_0^t |v(\tau)|^2d\tau }$ and thus
$\|e\|_{\mathcal{L}_2,[0,t)} \leq \sqrt{\frac{V(e(0))}{\varepsilon}} + \sqrt{\frac{\mu_w}{\varepsilon} }\|w\|_{\mathcal{L}_2,[0,t)} + \sqrt{\frac{\mu_v}{\varepsilon}} \|v\|_{\mathcal{L}_2,[0,t)}.$
Since $\lambda_\text{min}(P)|e'|^2\leq V(e')\leq\lambda_\text{max}(P)|e'|^2$, for any $e' \in \mathbb{R}^N$, we have $\sqrt{\frac{V(e(0))}{\varepsilon}} \leq \sqrt{\frac{\lambda_\text{max}(P)}{\varepsilon}}|e(0)|$. With this, we obtain $\|e\|_{\mathcal{L}_2,[0,t)} \leq \sqrt{\frac{\lambda_\text{max}(P)}{\varepsilon}}|e(0)| + \sqrt{\frac{\mu_w}{\varepsilon} }\|w\|_{\mathcal{L}_2,[0,t)} + \sqrt{\frac{\mu_v}{\varepsilon}} \|v\|_{\mathcal{L}_2,[0,t)}$, which implies the $\mathcal{L}_2$-stability of system (\ref{new_overallss}), (\ref{newobs}) as stated in Theorem \ref{T2}.

For the case when $w=0$ and $v=0$, we have $\langle \mathcal{r} V(e), Ae+Ew-L(h(x)-h(\hat{x}))-Lv+L(\Tilde{h}(x)-\Tilde{h}( \hat{x})) \rangle \leq -\varepsilon |e|^2$ and $\lambda_\text{min}(P)|e|^2\leq V(e)\leq\lambda_\text{max}(P)|e|^2$. Therefore,
we conclude the desired uniform global exponential stability property by following similar steps as in the proof of \cite[Theorem 4.10]{khalil2002nonlinear}.
$\hfill \blacksquare$

In addition to Assumptions \ref{A1} and \ref{A2}, Theorem \ref{T2} also requires (\ref{c2_th2}) to hold, which is a robustness property of observer (\ref{PierreObs}) with respect to $h_{\text{cor}}-h$. This condition is needed to ensure that the observer still provides satisfactory convergence properties when using $h_\text{cor}$ instead of $h$ to generate the estimated output, as in (\ref{newobs}). 
 \subsection{Polytopic approach}
 In this section, we directly synthesize an observer for system (\ref{new_overallss}) by following a similar approach as in \cite{Blondel_IFAC_Journal_2017, zemouche2008, DreefDonkersCDC2018}.
The proposed observer takes the same form as in (\ref{newobs}). However, in this section, $L$ is to be designed and not given by Assumption \ref{A2}.
 We consider the estimation error $e:= x-\hat{x}$, whose dynamics is the same as in (\ref{error}).

Another consequence of Assumption \ref{A1} is that the term  $h_\text{cor}(x)-h_\text{cor}(x')$ appearing in (\ref{error}) can be written as, for any $x, x' \in \mathbb{R}^N$
\begin{equation}
    h_\text{cor}(x)-h_\text{cor}(x')=C(x,x')(x-x'),
    \label{39}
\end{equation}
where $C(x,x'):=\sum_{i=1}^4 \Lambda_i(x,x')C_i$, with $C_{i} \in \mathbb{R}^N$ defined in (\ref{v2}), $\Lambda_i(x,x') \in [0,1]$ for $i \in \{1, 2, 3, 4\}$ and $\sum_{i=1}^4 \Lambda_i(x,x')=1$.
This means that 
$h_\text{cor}$ lies in a polytope defined by the vertices $C_i$ with $i \in \{1, 2, 3, 4\}$ in (\ref{v2}). Note the calculation of those vertices differ from those in (\ref{v1}) because of the change in the arguments of the OCVs in the output equation of the new model, see Section $\text{\rom{4}.C}$.

\begin{table*}
    \begin{equation}
    \begin{array}{ll}
        \tilde C_1 :=& \begin{pmatrix}
            \textbf{0}_{1 \times (N_\text{neg}-2)} & \left(-\frac{1}{c_\text{max}^\text{neg}}+\frac{K^{\text{neg}}_{N_\text{neg}}}{c_\text{max}^\text{neg}}\right)C_{\text{neg},1} & 
           -\overline{h}^\text{pos}_1C_{\text{pos},1}-h_1^*C_{\text{neg},1} & \hdots & 
            -\overline{h}^\text{pos}_{N_\text{pos}-1}C_{\text{pos},1}-h_{N_\text{pos}-1}^*C_{\text{neg},1} &

             \left(\frac{1}{c_\text{max}^\text{pos}}-\overline{h}^\text{pos}_{N_\text{pos}}- \frac{K^{\text{pos}}_{N_\text{pos}}}{c_\text{max}^\text{pos}} \right) C_{\text{pos},1}-h_{N_\text{pos}}^*C_{\text{neg},1}

        \end{pmatrix}   \\
        \tilde C_2 :=&\begin{pmatrix}
            \textbf{0}_{1 \times (N_\text{neg}-2)} & \left(-\frac{1}{c_\text{max}^\text{neg}}+\frac{K^{\text{neg}}_{N_\text{neg}}}{c_\text{max}^\text{neg}}\right)C_{\text{neg},1} & 
            -\overline{h}^\text{pos}_1C_{\text{pos},2}-h_1^*C_{\text{neg},1} & \hdots & 
            -\overline{h}^\text{pos}_{N_\text{pos}-1}C_{\text{pos},2}-h_{N_\text{pos}-1}^*C_{\text{neg},1} &

             \left(\frac{1}{c_\text{max}^\text{pos}}-\overline{h}^\text{pos}_{N_\text{pos}}- \frac{K^{\text{pos}}_{N_\text{pos}}}{c_\text{max}^\text{pos}} \right) C_{\text{pos},2}-h_{N_\text{pos}}^*C_{\text{neg},1}
        \end{pmatrix} \\
       \tilde C_3 :=&
        \begin{pmatrix}
            \textbf{0}_{1 \times (N_\text{neg}-2)} & \left(-\frac{1}{c_\text{max}^\text{neg}}+\frac{K^{\text{neg}}_{N_\text{neg}}}{c_\text{max}^\text{neg}}\right)C_{\text{neg},2} & 
            -\overline{h}^\text{pos}_1C_{\text{pos},1}-h_1^*C_{\text{neg},2} & \hdots & 
            -\overline{h}^\text{pos}_{N_\text{pos}-1}C_{\text{pos},1}-h_{N_\text{pos}-1}^*C_{\text{neg},2} &

             \left(\frac{1}{c_\text{max}^\text{pos}}-\overline{h}^\text{pos}_{N_\text{pos}}- \frac{K^{\text{pos}}_{N_\text{pos}}}{c_\text{max}^\text{pos}} \right) C_{\text{pos},1}-h_{N_\text{pos}}^*C_{\text{neg},2}

        \end{pmatrix} 
        \\
      \tilde  C_4 :=&
        \begin{pmatrix}
            \textbf{0}_{1 \times (N_\text{neg}-2)} & \left(-\frac{1}{c_\text{max}^\text{neg}}+\frac{K^{\text{neg}}_{N_\text{neg}}}{c_\text{max}^\text{neg}}\right)C_{\text{neg},2} & 
            -\overline{h}^\text{pos}_1C_{\text{pos},2}-h_1^*C_{\text{neg},2} & \hdots & 
            -\overline{h}^\text{pos}_{N_\text{pos}-1}C_{\text{pos},2}-h_{N_\text{pos}-1}^*C_{\text{neg},2} &

             \left(\frac{1}{c_\text{max}^\text{pos}}-\overline{h}^\text{pos}_{N_\text{pos}}- \frac{K^{\text{pos}}_{N_\text{pos}}}{c_\text{max}^\text{pos}} \right) C_{\text{pos},2}-h_{N_\text{pos}}^*C_{\text{neg},2}
        \end{pmatrix}. \\
        \end{array}\label{v1}\end{equation}
\vspace{0.2cm}
    \begin{equation}
    \begin{array}{ll}
        C_1 :=& \begin{pmatrix}
            \textbf{0}_{1 \times (N_\text{neg}-2)} & -\frac{K^{\text{neg}}_{N_\text{neg}}}{c_\text{max}^\text{neg}}C_{\text{neg},1} & 
           \overline{h}^\text{pos}_1C_{\text{pos},1}+h_1^*C_{\text{neg},1} & \hdots & 
            \overline{h}^\text{pos}_{N_\text{pos}-1}C_{\text{pos},1}+h_{N_\text{pos}-1}^*C_{\text{neg},1} &

             \left(\overline{h}^\text{pos}_{N_\text{pos}}+ \frac{K^{\text{pos}}_{N_\text{pos}}}{c_\text{max}^\text{pos}} \right) C_{\text{pos},1}+h_{N_\text{pos}}^*C_{\text{neg},1}

        \end{pmatrix}   \\
        C_2 :=&\begin{pmatrix}
            \textbf{0}_{1 \times (N_\text{neg}-2)} & -\frac{K^{\text{neg}}_{N_\text{neg}}}{c_\text{max}^\text{neg}}C_{\text{neg},1} & 
            \overline{h}^\text{pos}_1C_{\text{pos},2}+h_1^*C_{\text{neg},1} & \hdots & 
            \overline{h}^\text{pos}_{N_\text{pos}-1}C_{\text{pos},2}+h_{N_\text{pos}-1}^*C_{\text{neg},1} &

             \left(\overline{h}^\text{pos}_{N_\text{pos}}+ \frac{K^{\text{pos}}_{N_\text{pos}}}{c_\text{max}^\text{pos}} \right) C_{\text{pos},2}+h_{N_\text{pos}}^*C_{\text{neg},1}
        \end{pmatrix} \\
        C_3 :=&
        \begin{pmatrix}
            \textbf{0}_{1 \times (N_\text{neg}-2)} & -\frac{K^{\text{neg}}_{N_\text{neg}}}{c_\text{max}^\text{neg}}C_{\text{neg},2} & 
            \overline{h}^\text{pos}_1C_{\text{pos},1}+h_1^*C_{\text{neg},2} & \hdots & 
            \overline{h}^\text{pos}_{N_\text{pos}-1}C_{\text{pos},1}+h_{N_\text{pos}-1}^*C_{\text{neg},2} &

             \left(\overline{h}^\text{pos}_{N_\text{pos}}+ \frac{K^{\text{pos}}_{N_\text{pos}}}{c_\text{max}^\text{pos}} \right) C_{\text{pos},1}+h_{N_\text{pos}}^*C_{\text{neg},2}

        \end{pmatrix} 
        \\
        C_4 :=&
        \begin{pmatrix}
            \textbf{0}_{1 \times (N_\text{neg}-2)} & -\frac{K^{\text{neg}}_{N_\text{neg}}}{c_\text{max}^\text{neg}}C_{\text{neg},2} & 
            \overline{h}^\text{pos}_1C_{\text{pos},2}+h_1^*C_{\text{neg},2} & \hdots & 
            \overline{h}^\text{pos}_{N_\text{pos}-1}C_{\text{pos},2}+h_{N_\text{pos}-1}^*C_{\text{neg},2} &

             \left(\overline{h}^\text{pos}_{N_\text{pos}}+ \frac{K^{\text{pos}}_{N_\text{pos}}}{c_\text{max}^\text{pos}} \right) C_{\text{pos},2}+h_{N_\text{pos}}^*C_{\text{neg},2}
        \end{pmatrix}. \\
        \end{array}\label{v2}\end{equation}
\end{table*}

In view of (\ref{39}), the estimation error dynamics can be written as
\begin{equation}
     \Dot{e}=(A-LC(x,\hat{x}))e+Ew-Lv.
\end{equation}
The next theorem provides a sufficient condition to design gain $L \in \mathbb{R}^{N}$ under which $e=0$ is globally exponentially stable in absence of noise $v$ and disturbance $w$, and satisfies $\mathcal{L}_2$-stability properties when the latter are non-zero.

\begin{thm}
  Suppose Assumption \ref{A1} holds and there exist $\epsilon, \mu_w, \mu_v \in \mathbb{R}_{>0}$, $L \in \mathbb{R}^{N}$ and $P \in \mathbb{R}^{N \times N}$
symmetric and
positive definite such that for any $i \in \{1, . . . , 4 \}$
\begin{equation}
\begin{pmatrix}
  \mathcal{H}_i+\epsilon I_N & PE & -PL\\
* & -\mu_wI_{n_w} & 0 \\
* & * &  -\mu_vI_{n_v}
\end{pmatrix}
\leq 0,
\label{LMI}
\end{equation}
with $\mathcal{H}_i
:= (A-LC_i)^T P + P(A-LC_i)$ then system (\ref{new_overallss}), (\ref{newobs}) is
$\mathcal{L}_2$-stable from ($w,v$) to $e$ with gain less or equal to $\sqrt{\frac{\mu_w}{\epsilon}}$ and $\sqrt{\frac{\mu_v}{\epsilon}}$, respectively, and $e=0$ uniformly globally exponentially stable when $w$=0 and $v$=0.
$\hfill \Box$
\label{T3}
\end{thm}

The proof of Theorem \ref{T3} follows the same steps as in \cite[Theorem 1]{Blondel_IFAC_Journal_2017} and is therefore omitted. Theorem \ref{T3} means that we can design $L$ to ensure the exponential convergence of the state estimate to the true state whenever (\ref{LMI}) is verified. 
 The matrix inequality in (\ref{LMI}) is not linear, however it becomes linear
after a standard change of variables, namely $W = P L$. Condition (\ref{LMI}) can be easily tested given the model parameters as done in Section VI.B. Also, the order reduction performed in Section \text{\rom{4}}.B appear to be essential for its feasibility, see also \cite{9920541} where a similar condition is imposed for a different battery model.
\subsection{Corrected estimated concentrations}

The observers in Sections $\text{\rom{5}.B}$ and $\text{\rom{5}.C}$ generate estimated lithium concentrations, which can be corrected along with $\hat c_{\text{neg},1}$ so that they asymptotically match the concentrations of the PDEs in (\ref{PDE}) for constant input currents as seen in Section III. We note that from $\hat{x}$, which represents the concatenation of the estimated concentrations generated by the chosen observer $\hat{x}=(\hat{c}_{\text{neg},2},\hdots,\hat{c}_{\text{neg},{N_\text{neg}}},\hat{c}_{\text{pos},1},\hdots,\hat{c}_{\text{pos},{N_\text{pos}}})$, we can retrieve $\hat{c}_{\text{neg},1}$ by replacing the concentrations in (\ref{model-red}) by their estimates. We denote in the following $c_{s, (1)}$ the concentrations generated by the PDEs in (\ref{PDE}) as in Section III and $\hat c_s$ the estimated concentrations of electrode $s$, with $s \in \{\text{pos}, \text{neg}\}$, given by (\ref{newobs}) where $L$ is obtained either from Assumption \ref{A2} or by verifying the conditions of Theorem \ref{T3}. 

We denote the corrected estimated concentrations as $\hat{c}_{s,\text{cor}}$, which are given by, for $j \in \{1,\hdots,N_s\}$ and $s\in\{\text{neg},\text{pos}\}$,
\begin{equation}
    \hat{c}_\text{s,cor,j}:=\hat{c}_{s,\text{mean}}-K^{s}_j(\hat{c}_{s,\text{mean}}-\hat{c}_{s,j}),
    \label{corr-est-conc}
\end{equation}
where $\hat{c}_{s,\text{mean}}:=\frac{1}{V_s}\sum_{i=1}^{N_s}V^s_i\hat{c}_{s,i}$ and $K^{s}_j$ defined in (\ref{coeff_corr}).

The next theorem states an asymptotic property of the error between the corrected estimated concentrations and the concentrations generated by the original PDEs in (\ref{PDE}) when time tends to infinity in absence of disturbances and noises for constant inputs.
\begin{thm}
    Consider system (\ref{new_overallss}) and its corresponding observer (\ref{newobs}) and suppose $e=x-\hat{x}=0$ is globally exponentially stable when $w=0$ and $v=0$. Then, for any constant input current $I_\text{cell}$ and any $c_0\in\R_{\geq 0}$ such that $c_{s,(1)}(r,0)=c_0$ for all $r \in [0,R_s]$, any corresponding solution $\hat{x}$ to (\ref{newobs}) and $c_{s,(1)}$ to (\ref{PDE}) satisfy
\begin{equation}
    (\hat{c}_\text{cor}-c_{(1)})_\infty=0,
    \label{eq_theorem4}
\end{equation}
    where $\hat{c}_\text{cor}:=(\hat{c}_{\text{neg,cor},1}, \hat{c}_{\text{neg,cor},2},\hdots,\hat{c}_{\text{neg,cor},{N_\text{neg}}},\hat{c}_{\text{pos,cor},1},\hdots,$ $\hat{c}_{\text{pos,cor},{N_\text{pos}}})$ is the vector of the corrected estimated concentrations and $c_{(1)}:=(c_{\text{neg},(1)}(r_1^\text{neg},\cdot),
    c_{\text{neg},(1)}(r_2^\text{neg},\cdot), \hdots, c_{\text{neg},(1)}(r_{N_\text{neg}}^\text{neg},\cdot),c_{\text{pos},(1)}(r_1^\text{pos},\cdot)$,
    $\hdots, c_{\text{pos},(1)}(r_{N_\text{pos}}^\text{pos},\cdot))$. $\hfill \Box$
    \label{T4}
\end{thm}
\textbf{Proof.} 
We consider system (\ref{new_overallss}), (\ref{newobs}) and suppose $e=x-\hat{x}=0$ is globally exponentially stable when $w=0$ and $v=0$. Let $I_\text{cell}$ be a constant input, $c_{s,(1)}(\cdot,0)=c_0$ with $c_0\in\R_{\geq 0}$, $c_{s,(1)}$ be the corresponding solution to (\ref{PDE}) and $(x,\hat{x})$ be a corresponding solution to (\ref{new_overallss}), (\ref{newobs}). We thus have $(\hat{x}-x)_\infty=0$ and $(\hat c_{\text{neg},1}- c_{\text{neg},1})_\infty=0$. Hence, in view of (\ref{conc_corr}) and (\ref{corr-est-conc})
\begin{equation}
    (\hat c_\text{cor} - c_\text{cor})_\infty=0,
    \label{46}
\end{equation}
where ${c}_\text{cor}:=({c}_{\text{neg,cor},1},{c}_{\text{neg,cor},2},\hdots,{c}_{\text{neg,cor},{N_\text{neg}}},{c}_{\text{pos,cor},1},\hdots$,
${c}_{\text{pos,cor},{N_\text{pos}}})$ and $\hat{c}_\text{cor}:=(\hat{c}_{\text{neg,cor},1},\hat{c}_{\text{neg,cor},2},\hdots,\hat{c}_{\text{neg,cor},{N_\text{neg}}},$
$\hat{c}_{\text{pos,cor},1},\hdots,\hat{c}_{\text{pos,cor},{N_\text{pos}}})$.

In view of Theorem \ref{T1}, we have that $(c_{s,\text{cor},j}-c_{s,(1)}(r_j^s, \cdot))_\infty=0$ for any $j \in \{1,\hdots,N_s\}$ and $s\in\{\text{neg},\text{pos}\}$. Therefore, we obtain 
\begin{equation}
    (c_\text{cor}-c_{(1)})_\infty=0,
    \label{47}
\end{equation}
where $c_{(1)}:=(c_{\text{neg},(1)}(r_1^\text{neg}, \cdot),c_{\text{neg},(1)}(r_2^\text{neg}, \cdot), \hdots, c_{\text{neg},(1)}(r_{N_\text{neg}}^\text{neg}, \cdot)$,
$c_{\text{pos},(1)}(r_1^\text{pos}, \cdot), \hdots$,
$c_{\text{pos},(1)}(r_{N_\text{pos}}^\text{pos}), \cdot)$.
From (\ref{46}) and (\ref{47}), we derive $(\hat c_\text{cor} - c_{(1)})_\infty=0$ as in (\ref{eq_theorem4}), which completes the proof.$\hfill \blacksquare$

Theorem \ref{T4} implies that, under the conditions of Theorems \ref{T2} or \ref{T3}, the estimated corrected lithium concentrations asymptotically match the lithium concentrations of the PDEs in (\ref{PDE}) in absence of noise and disturbance provided that a constant input current is applied and $c_{s,(1)}(r,0)$ is uniform for all $r \in [0,R_s]$.

\section{Numerical case study}
In this section, we illustrate numerically the benefits of the new, corrected model and the associated state estimation scheme. For this purpose, we consider the infinite-dimensional model in \cite{RAEL2013112} as a reference model and we first compare it with the model in \cite{Blondel_TCST_Journal_2019} without correction and model (\ref{new_overallss}) with correction considering a uniform volume discretization method (i.e, $V^s_i=\frac{V_s}{N_s}$ for $i \in \{1,\hdots,N_s\}$) as done in \cite{Blondel_TCST_Journal_2019} (Section ${\text{\rom{6}.A}}$). Next, we synthesize observers based on the polytopic approach for the model in \cite{Blondel_TCST_Journal_2019} and model (\ref{new_overallss}) using Theorem \ref{T3} (the conditions of Theorem \ref{T2} were not satisfied) and then we compare the obtained estimated variables (Section ${\text{\rom{6}.B}}$).

\begin{figure}[thpb]
    \centering
    \includegraphics[scale=0.319]{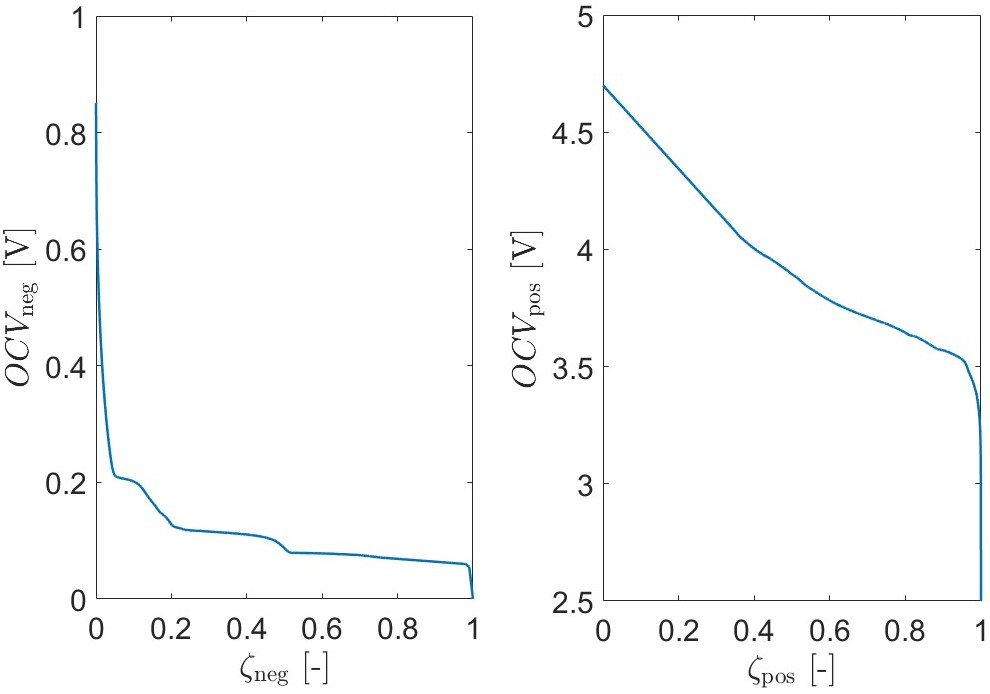}
    \vspace{-0.2cm}
    \caption{OCVs curves.}
    \label{OCV}
\end{figure}

\begin{figure}[thpb]
    \centering
    \includegraphics[scale=0.33]{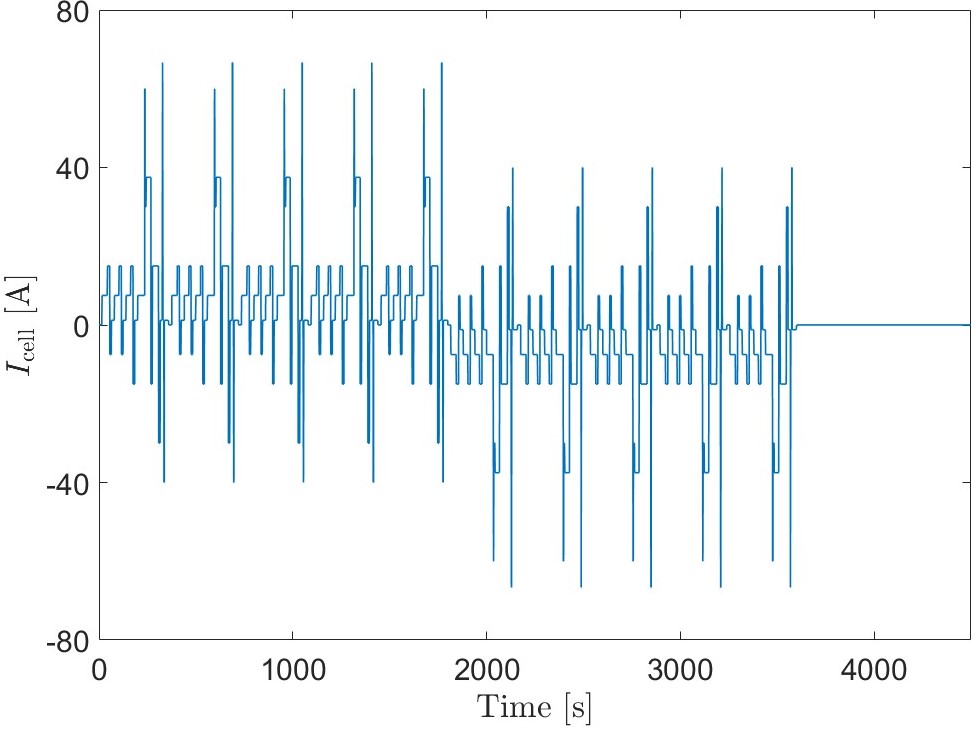}
    \vspace{-0.2cm}
    \caption{Input current profile.}
    \label{current}
\end{figure}

We simulate model (\ref{new_overallss}) and the model in \cite{Blondel_TCST_Journal_2019} with the parameters values given in Table \ref{param_table}. We assume that the particles in each electrode are discretized into 4 samples of uniform volume. Consequently, $N_\text{neg}=N_\text{pos}=4$ and $N=4+4-1=7$; recall that the concentration at the center of the negative is electrode can be removed as explained in Section $\text{\rom{4}.B}$. No measurement noise or perturbation are considered when simulating the two models. The OCVs curves are given in Figure \ref{OCV}. These OCVs are extrapolated using first order approximations on $(-\infty,0]$ and $[1,\infty)$, respectively, which implies that the OCVs are globally Lipschitz on $\mathbb{R}$ and thus satisfy Assumption \ref{A1} as explained in Section V.A. In particular, Assumption \ref{A1} holds with $C_{\text{neg},1}=-75.2267, C_{\text{neg},2}=-0.0067, C_{\text{pos},1}=-1266.7, C_{\text{pos},2}=-0.2667$. The used input current $u$ is a Plug-in Hybrid Electrical Vehicle (PHEV) discharge current on the time interval $[0,1800]$, a PHEV charge current on the time interval $[1800,3600]$ and $0$ on $[3600,4500]$, as illustrated in Figure \ref{current}. The current profile is thus rapidly varying on [0,3600], during which we will see that improvements are obtained with the corrections presented in Section III. We initialize both models at equilibrium, meaning
that all the initial concentrations within the same
electrode are equal, with a SOC equal to
100$\%$. The SOC is defined by for $s \in \{\text{pos, \text{neg}}\}$ 
     \begin{equation}
         SOC_s:=100\frac{c_{s,\text{mean}}-c_0^s}{c_{100}^s-c_0^s},
         \label{soc_eq}
     \end{equation}
where $c_0^s$, $c_{100}^s$ are the lithium concentration of electrode $s$ at SOC equal $0 \%$ and 100$\%$, respectively, see Table \ref{param_table}. We note that $c_{s,\text{mean}}:=\frac{1}{N_s}\sum_{i=1}^{N_s}c_{s,i}$ (recall that the discretization method we chose is a uniform volume discretization). Given that $SOC_\text{pos}$ and $SOC_\text{neg}$ are equal, we use the notation $SOC$ instead.
As for the reference model, it is obtained by solving the PDEs of (\ref{PDE}) using a finite elements method and simulated with the parameters of Table \ref{param_table}, see \cite{RAEL2013112} for details.

\begin{figure}[thpb]
    \centering
    \includegraphics[scale=0.319]{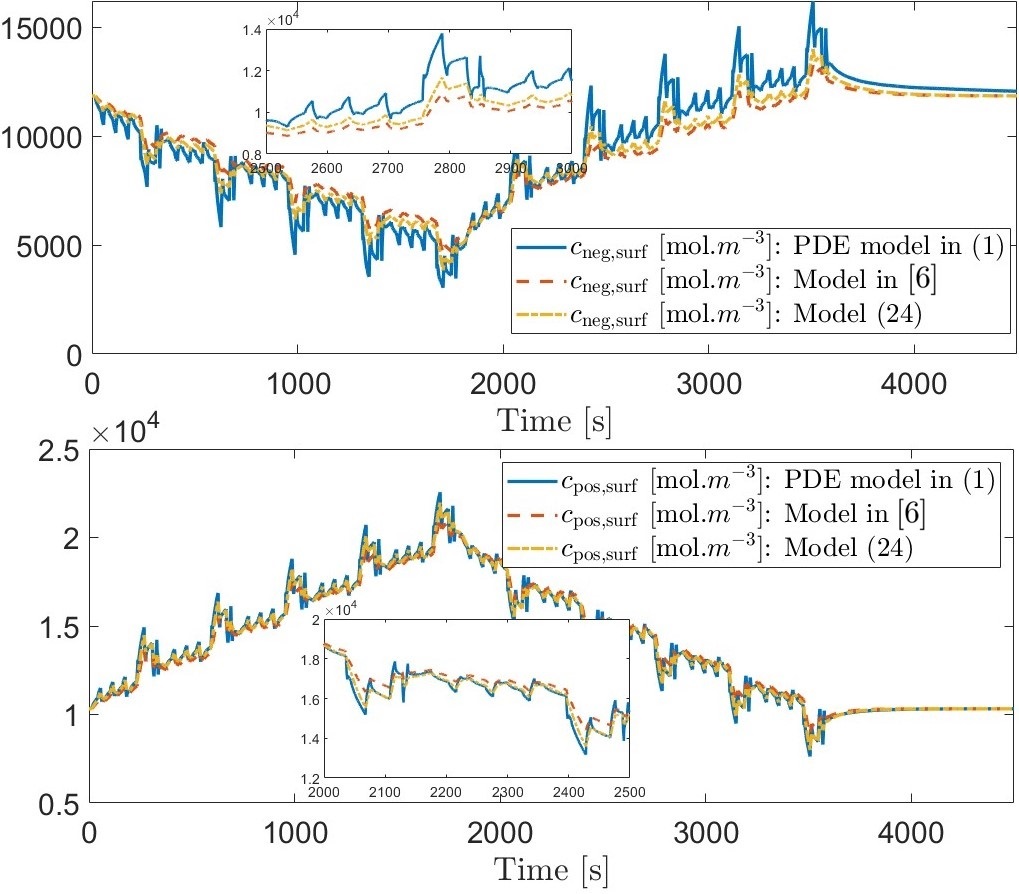}
    \vspace{-0.2cm}
    \caption{Surface concentrations.}
    \label{surfaceconc1}
\end{figure}
\begin{figure}[thpb]
    \centering
    \includegraphics[scale=0.185]{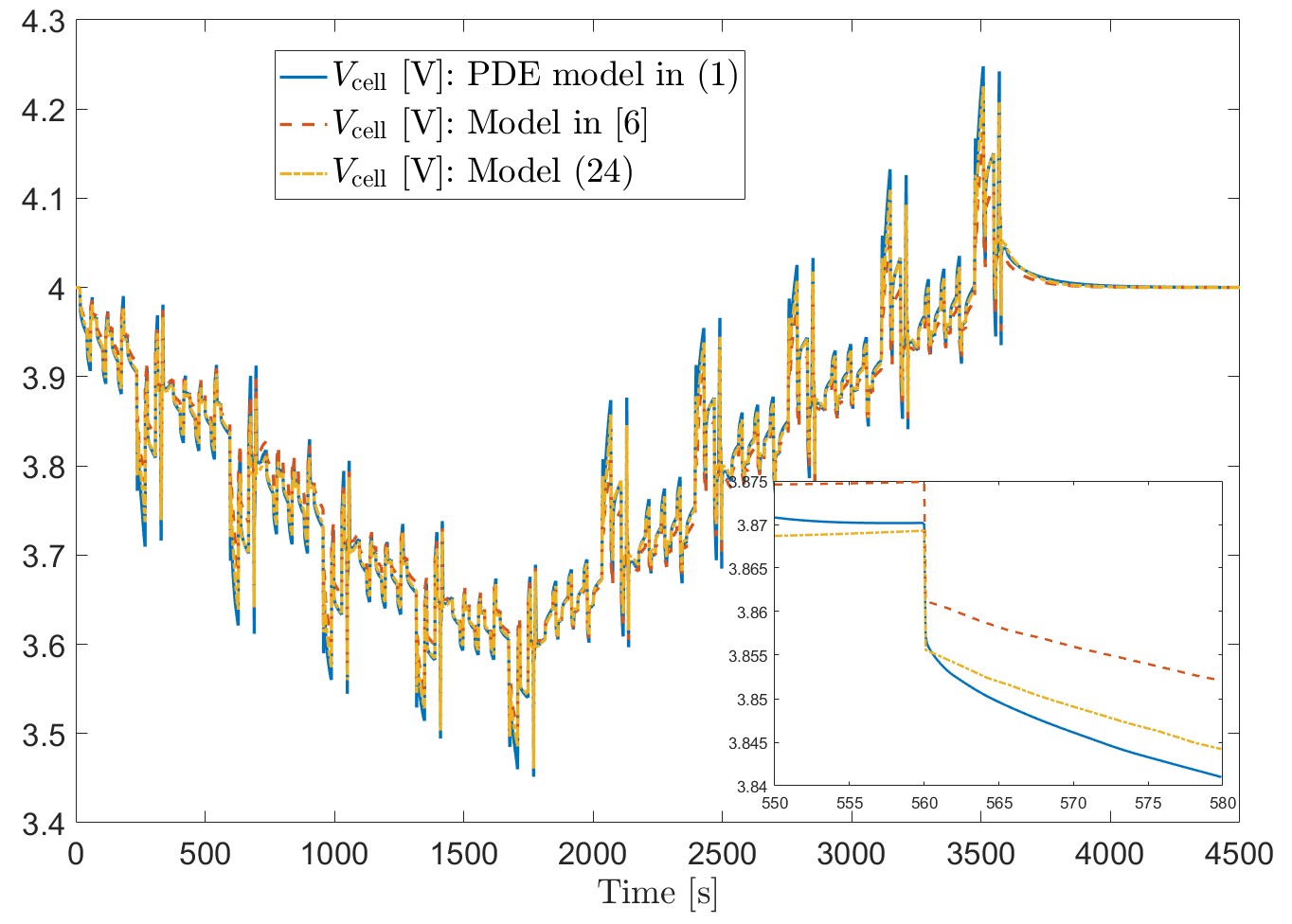}
    \vspace{-0.2cm}
    \caption{Output voltage.}
    \label{vcell}
\end{figure}

\subsection{Models comparison}
We have compared the surface concentrations $c_{s,\text{surf}}$ and the output voltages $V_\text{cell}$ generated by the model in \cite{Blondel_TCST_Journal_2019} and model (\ref{new_overallss}), with those generated by the reference infinite-dimensional model in Figures \ref{surfaceconc1} and \ref{vcell}, respectively, over the whole interval of time [0,4500] as well as [0,3600] where the current is rapidly varying. We see that, in all cases, the proposed corrections allow improving the accuracy of the data even over short time horizons and when the input is rapidly changing. The considered surface concentrations of the reference infinite-dimensional model are those taken at the electrodes/separator interfaces. To quantify this improvement, we have computed the mean absolute error (MAE) and the root mean square error (RMSE) of the voltage error $e_{V_\text{cell}}$ between $V_\text{cell}$ generated by (\ref{PDE}) and $V_\text{cell}$ generated by the model in \cite{Blondel_TCST_Journal_2019} and by (\ref{new_overallss}), respectively. We similarly computed the normalized surface concentrations error $\textbf{e}_{c_{s,\text{surf}}}$ between the surface concentrations generated by (\ref{PDE}) and those generated by the model in \cite{Blondel_TCST_Journal_2019} and model (\ref{new_overallss}), respectively. The results are given in
Table \ref{errortable0}. Model (\ref{new_overallss}) is more accurate than the one in \cite{Blondel_TCST_Journal_2019}. In particular, we see improvements of the order of 50$\%$ for the output voltage, $50\%$ for the surface concentration of the positive electrode and $30\%$ for the surface concentration of the negative electrode on the interval [0,4500] and particularly on [0,3600], when the current is rapidly varying.
\begin{table}[h]
    \centering
    \begin{tabular}[t]{l|llll}
         & MAE & RMSE & MAE & RMSE  \\
         & [0,4500]& [0,4500] &  [0,3600] & [0,3600] \\
         \hline 
       $e_{V_\text{cell}}$: model in \cite{Blondel_TCST_Journal_2019} [mV] &  12.07 & 17.76 & 14.57 & 19.80\\
    $e_{V_\text{cell}}$: model (\ref{new_overallss}) [mV]&\textbf{5.07}  & \textbf{8.28} & \textbf{6.09}&\textbf{9.23}\\ Improvement [\%]   & \textbf{58.0} & \textbf{53.4} & \textbf{58.2}& \textbf{53.4} \\
     \hline 
    $\textbf{e}_{c_{\text{pos},\text{surf}}}$: model in \cite{Blondel_TCST_Journal_2019} [\%] &  2.05 & 3.01 & 2.48 & 3.36 \\
    $\textbf{e}_{c_{\text{pos},\text{surf}}}$: model (\ref{new_overallss}) [\%] & \textbf{0.95}  & \textbf{1.53} & \textbf{1.15}&\textbf{1.71} \\ Improvement [\%]   & \textbf{53.7} & \textbf{49.2} & \textbf{53.6}& \textbf{49.1}\\
    \hline 
        $\textbf{e}_{c_{\text{neg},\text{surf}}}$: model in \cite{Blondel_TCST_Journal_2019} [\%]  & 8.51 & 11.79 & 9.93&13.10  \\
    $\textbf{e}_{c_{\text{neg},\text{surf}}}$: model (\ref{new_overallss}) [\%] &\textbf{5.48}   & \textbf{7.79}  & \textbf{6.24} & \textbf{8.62}\\ Improvement [\%]   & \textbf{35.6 }& \textbf{33.9} & \textbf{37.2}&\textbf{34.2}\\
    
   \end{tabular}%
 \vspace{0.4cm}
    \caption{MAE and RMSE of the output voltage and the surface concentrations errors given by the model in \cite{Blondel_TCST_Journal_2019} and model (\ref{new_overallss}).}
      \label{errortable0}
      \vspace{-0.6cm}
\end{table}

\subsection{State estimation}
We have designed observer (\ref{newobs}) for system (\ref{new_overallss}) by applying\footnote{We have not been able to ensure condition (\ref{c2_th2}) using the observer design technique in \cite{Blondel_IFAC_Journal_2017}. Nevertheless, this is not an issue, as, again, an observer for the corrected model can be synthesized using Theorem \ref{T3}.} Theorem \ref{T3}, as (\ref{LMI}) holds for the considered parameter values taking $E=B$ with  $L=10^4$(3.2387,3.5432,3.3896,-5.0388,-5.7421,-5.3310,-5.433750), $P=10^{-9}\scriptsize{\begin{pmatrix}
    0.0137 & 0.0258 & 0.0329 & 0.0066 & 0.0107 & 0.0135 & 0.0149 \\

    0.0258 & 0.0550 & 0.0797 & 0.0127 & 0.0220 & 0.0304 & 0.0361 \\

    0.0329 & 0.0797 & 0.1485 & 0.0179 & 0.0312 & 0.0474 & 0.0681 \\

    0.0066 & 0.0127 & 0.0179 & 0.0136 & 0.0095 & 0.0039 & -0.0031 \\

    0.0107 & 0.0220 & 0.0312 & 0.0095 & 0.0117 & 0.0115 & 0.0077 \\

    0.0135 & 0.0304 & 0.0474 & 0.0039 & 0.0115 & 0.0190 & 0.0231 \\

    0.0149 & 0.0361 & 0.0681 & -0.0031 & 0.0077 & 0.0231 & 0.0471 \\
    
\end{pmatrix}}$, $\epsilon=1.17^*10^{-22}$, $\mu_v=7.9784$ and $\mu_w=1.0486$. For the sake of comparison, we have designed an observer for the non-corrected model (\ref{oldss}) using the technique in \cite{Blondel_IFAC_Journal_2017}, which ensures Assumption \ref{A2} holds with the same gain $L$. The only difference between the two observers is the output equation used to synthesize them.
We have initialized both observers such that all estimated concentrations within the same particle are equal and correspond to a SOC estimate, denoted $\widehat{SOC}$, of $0 \%$. We note that $\widehat{SOC}$ is obtained by replacing $c_{s,\text{mean}}$ in (\ref{soc_eq}) by its estimate $\hat{c}_{s,\text{mean}}$. In practical applications, the observers only know a biased version of the input current $u$ and the output voltage $y$. This bias is due to the precision
of the sensors and their conditioning. Therefore, we introduce $I_\text{cell, biased}$, the input $u$ known by the observer, which is given by $I_\text{cell, biased}:=I_\text{cell}+3\sin(2000\pi t)$. As for the output voltage $y$ of the observers, it becomes the output voltage generated by the
infinite-dimensional reference model in \cite{RAEL2013112} in addition to a bias given by $0.05\sin(200 \pi t)$.

\begin{figure}[thpb]
    \centering
    \includegraphics[scale=0.215]{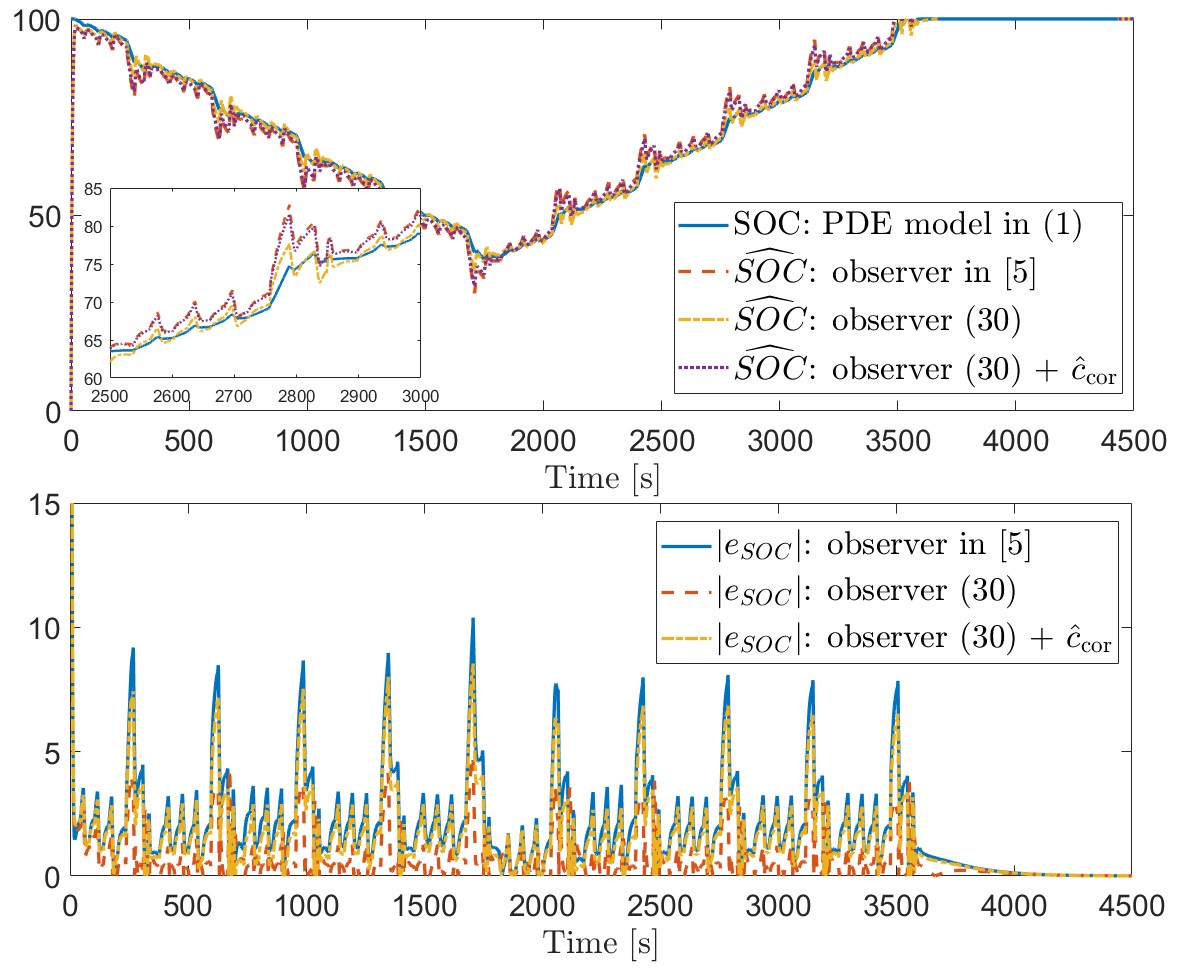}
    \vspace{-0.2cm}
    \caption{$SOC$ and $\widehat{SOC}$ generated by the observer in \cite{Blondel_IFAC_Journal_2017}, observer (\ref{newobs}) and observer (\ref{newobs}) with $\hat{c}_\text{cor}$
(top) and norm of the estimation errors $e_{SOC} := SOC-\widehat{SOC}$ (bottom).}
    \label{soc}
\end{figure}

Figure \ref{soc} reports the actual SOC given by the infinite-dimensional model and the estimated ones, as well as the corresponding norm of the estimation errors on the SOC $e_{SOC}=SOC-\widehat{SOC}$ obtained with the observer in \cite{Blondel_IFAC_Journal_2017}, observer (\ref{newobs}) and observer (\ref{newobs}) with the correction of its estimated concentrations $\hat c:=(\hat c_{\text{neg},1}, \hat x)$ as done in Theorem \ref{T4}.  We see that observer (\ref{newobs}) based on the corrected model (\ref{new_overallss}) provides a more accurate SOC. This improvement is quantified by computing the MAE and the RMSE
of the SOC estimation errors $e_{SOC}$ for the observer in \cite{Blondel_IFAC_Journal_2017}, observer (\ref{newobs}) and observer (\ref{newobs}) with $\hat{c}_\text{cor}$,
respectively, averaged over 20 simulations for initial SOC
estimates ranging from $\{0\%, 5\%,\hdots, 100\%\}$ and for different gains values $L$, $10L$ and $L/10$. The results are given in
Table \ref{errortable}, where the percentages in parenthesis represent the improvement of the associated observer compared to the observer in \cite{Blondel_IFAC_Journal_2017}. We see that observer (\ref{newobs}) generates more accurate results in terms of SOC estimation and this improvement is of the order of the percent, which is significant for lithium-ion batteries. 

We have also computed in Table \ref{errortable} the average MAE and the average RMSE, over the same 20 scenarios and for the same 3 gains, of the normalized estimated concentrations error $\textbf{e}_{c_s}$ for $s \in \{\text{pos}, \text{neg}\}$, for the observer in \cite{Blondel_IFAC_Journal_2017}, for observer (\ref{newobs}) and for observer (\ref{newobs}) followed by the correction of the estimated concentrations ($\hat{c}_\text{cor}$).
The normalized estimated concentrations error $\textbf{e}_{c_s}$ is defined as follows
\begin{equation}
    \textbf{e}_{c_s}:=\frac{|c_s-\hat c_s|}{|c_s|},
\end{equation}
where $c_s := (c_s(r^s_1, \cdot), c_s(r^s_2, \cdot), c_s(r^s_3, \cdot), c_s(r^s_4, \cdot))$ is the vector
of concentrations generated by (\ref{PDE}) at the electrodes/separator
interfaces and $\hat c_s$ is the vector of estimated concentrations
generated by the chosen observer.
Table \ref{errortable} reports that more accurate estimated concentrations are obtained as a result of the correction of the estimated concentrations.

\begin{table}[h]
\footnotesize
    \centering

    \begin{tabular}{l|l|lll}
       &  & MAE & RMSE  &   \\ 
         \hline
     \multirow{9}{*}{$L$}& $e_{SOC}$: observer in \cite{Blondel_IFAC_Journal_2017} [\%]&  1.92 & 2.76  \\
      &$e_{SOC}$: observer (\ref{newobs}) [\%]& \textbf{0.81 (57.8)} & \textbf{1.35 (51.1)} \\
      &$e_{SOC}$: observer (\ref{newobs}) + $\hat{c}_\text{cor}$ [\%]& {1.58} & {2.29} \\
      \cline{2-4}
     & $\textbf{e}_{{c}_\text{pos}}$: observer in \cite{Blondel_IFAC_Journal_2017} [\%]& 1.16 & 1.87\\
     & $\textbf{e}_{{c}_\text{pos}}$: observer (\ref{newobs}) [\%]& 1.79 & 2.35 \\
     & $\textbf{e}_{{c}_\text{pos}}$: observer (\ref{newobs}) + $\hat{c}_\text{cor}$ [\%]& \textbf{1.02 (12.1)} & \textbf{1.65 (11.8)} \\
  \cline{2-4}
      & $\textbf{e}_{{c}_\text{neg}}$: observer in \cite{Blondel_IFAC_Journal_2017} [\%]& 5.82 & 6.5\\
       &     $\textbf{e}_{{c}_\text{neg}}$: observer (\ref{newobs}) [\%]& 7.28 & 8.18 \\
       & $\textbf{e}_{{c}_\text{neg}}$: observer (\ref{newobs}) + $\hat{c}_\text{cor}$ [\%]& \textbf{4.99 (14.3)} & \textbf{5.57 (14.3)} \\
       \hline
          \multirow{9}{*}{$10L$}& $e_{SOC}$: observer in \cite{Blondel_IFAC_Journal_2017} [\%]&  1.95 & 2.81 \\
      &$e_{SOC}$: observer (\ref{newobs}) [\%]& \textbf{0.88 (54.9)} & \textbf{1.43 (49.1)} \\
      &$e_{SOC}$: observer (\ref{newobs}) + $\hat{c}_\text{cor}$ [\%]& {1.61} & {2.33} \\
     \cline{2-4}
     & $\textbf{e}_{{c}_\text{pos}}$: observer in \cite{Blondel_IFAC_Journal_2017} [\%]& 1.26 & 1.90\\
     & $\textbf{e}_{{c}_\text{pos}}$: observer (\ref{newobs}) [\%]& 1.89 & 2.40 \\
     & $\textbf{e}_{{c}_\text{pos}}$: observer (\ref{newobs}) + $\hat{c}_\text{cor}$ [\%]& \textbf{1.13 (10.3)} & \textbf{1.66 (12.6)} \\
  \cline{2-4}
      & $\textbf{e}_{{c}_\text{neg}}$: observer in \cite{Blondel_IFAC_Journal_2017} [\%]& 5.81 & 6.49\\
       &     $\textbf{e}_{{c}_\text{neg}}$: observer (\ref{newobs}) [\%]& 7.28 & 8.15 \\
       & $\textbf{e}_{{c}_\text{neg}}$: observer (\ref{newobs}) + $\hat{c}_\text{cor}$ [\%]& \textbf{4.97 (14.5)} & \textbf{5.57 (14.2)} \\
       \hline
          \multirow{9}{*}{$L/10$}& $e_{SOC}$: observer in \cite{Blondel_IFAC_Journal_2017} [\%]&  1.9 & 2.89  \\
      &$e_{SOC}$: observer (\ref{newobs}) [\%]& \textbf{0.73 (61.6)} & \textbf{1.63 (43.6)} \\
      &$e_{SOC}$: observer (\ref{newobs}) + $\hat{c}_\text{cor}$ [\%]& {1.59} & {2.53} \\
      \cline{2-4}
     & $\textbf{e}_{{c}_\text{pos}}$: observer in \cite{Blondel_IFAC_Journal_2017} [\%]& 1.12 & 2.42\\
     & $\textbf{e}_{{c}_\text{pos}}$: observer (\ref{newobs}) [\%]& 1.82 & 2.86 \\
     & $\textbf{e}_{{c}_\text{pos}}$: observer (\ref{newobs}) + $\hat{c}_\text{cor}$ [\%]& \textbf{1.05 (6.25)} & \textbf{2.34 (3.3)} \\
  \cline{2-4}
      & $\textbf{e}_{{c}_\text{neg}}$: observer in \cite{Blondel_IFAC_Journal_2017} [\%]& 5.82 & 6.54\\
       &     $\textbf{e}_{{c}_\text{neg}}$: observer (\ref{newobs}) [\%]& 7.31 & 8.22 \\
       & $\textbf{e}_{{c}_\text{neg}}$: observer (\ref{newobs}) + $\hat{c}_\text{cor}$ [\%]& \textbf{5.02 (13.7)} & \textbf{5.65 (13.6)} \\
      
   \end{tabular}
   \vspace{0.4cm}
    \caption{Average MAE and RMSE over 20 simulations of the SOC estimation errors and of the estimated concentrations error for different gain values. The values in parenthesis represent the percentage of improvement with respect to the observer in \cite{Blondel_IFAC_Journal_2017}.}
    \label{errortable}
    \vspace{-0.6cm}
\end{table}

\section{Experimental validation}
\begin{figure}[thpb]
    \centering
    \includegraphics[scale=0.2]{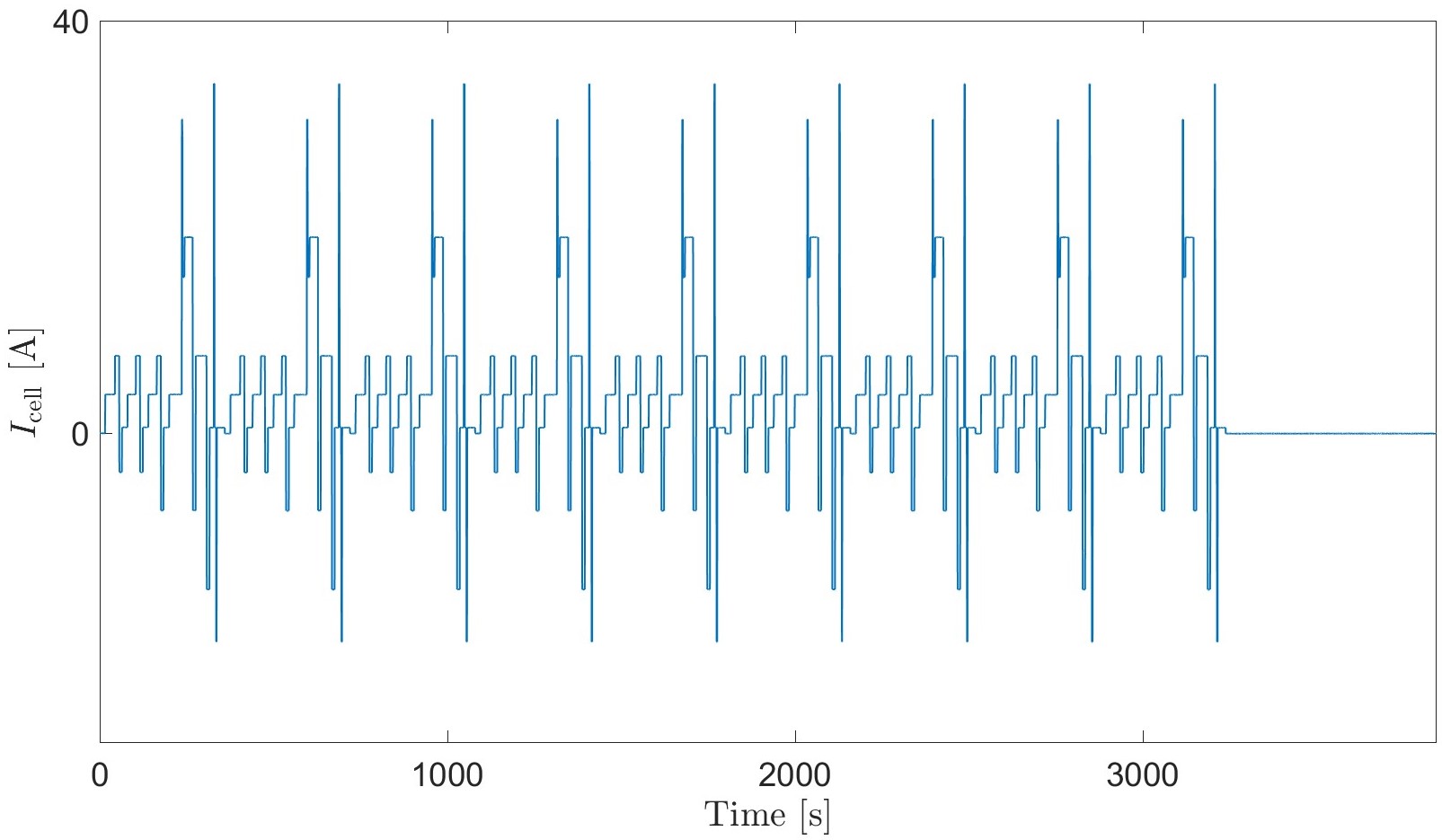}
    \vspace{-0.2cm}
    \caption{Measured input current.}
    \label{expcurrent}
\end{figure}

\begin{figure}[thpb]
    \centering
    \includegraphics[scale=0.2]{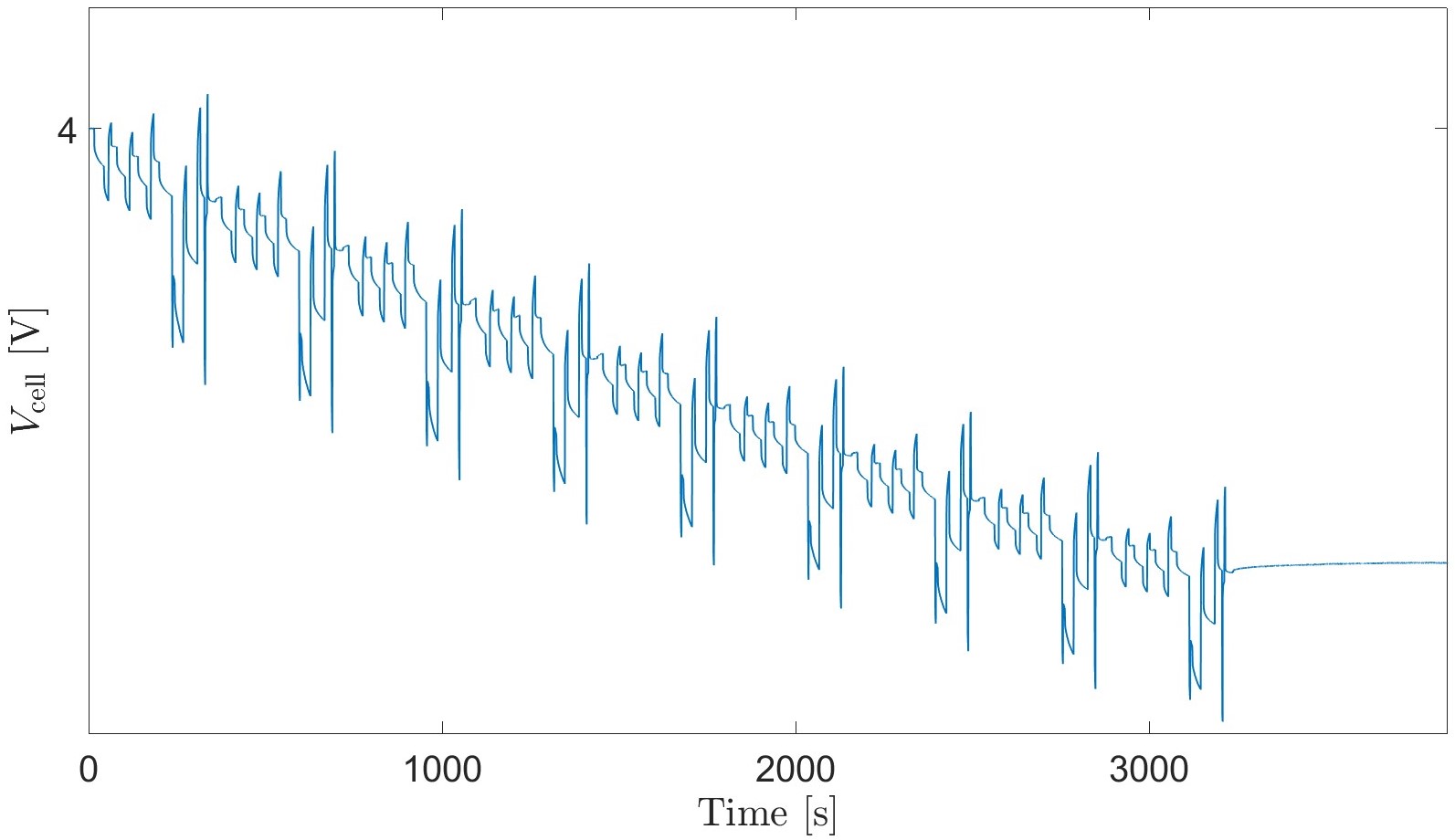}
    \vspace{-0.2cm}
    \caption{Measured output voltage.}
    \label{expvol}
\end{figure}

In this section, an experimental validation of the obtained results is carried out on a 6 Ah lithium-ion battery cell, fitted with a graphite negative electrode and a NCA positive electrode. The cell parameters required for model computation are detailed in Table \ref{param_table}. Some of these parameters are taken from \cite{Smith2006JournalofPowerSources} (electrode thicknesses, particle radius, volume fractions, electronic and ionic conductivities), the others have been estimated by experimental characterizations. The cell current and voltage are measured using sensors. We thus obtain the current input shown in Figure \ref{expcurrent}, which represents a PHEV discharge current on the interval [0,3240] and 0 on [3240,3844], and the measured output voltage shown in Figure \ref{expvol}.

We first compare the output voltages of both model \cite{Blondel_TCST_Journal_2019} without correction and model (\ref{new_overallss}) with correction considering a uniform volume discretization as in \cite{Blondel_TCST_Journal_2019}. For this purpose, we consider the same OCV curves as in Figure \ref{OCV} and we feed both models with a corrected version of the measured current of Figure \ref{expcurrent}. This current correction is done via a multiplicative gain equal to $1.035$, which is chosen such that the measured output voltage and the output voltage obtained by the infinite-dimensional model in \cite{RAEL2013112} match in the terminal steady state. We have calculated the MAE and RMSE of the voltage error $e_{V_\text{cell}}$ between the measured output voltage and the voltage generated by the model in \cite{Blondel_TCST_Journal_2019} and model (\ref{new_overallss}), respectively. We have obtained the results of Table \ref{experrortable0}. We see that the output voltage of the corrected model is about 25$\%$ more accurate  than the model in \cite{Blondel_TCST_Journal_2019}.  

\begin{table}[h]
    \centering
    \begin{tabular}[t]{l|ll|ll}
         & MAE & RMSE  & MAE & RMSE  \\
         & [0,3844] & [0,3844] & [0,3240] &[0,3240]\\
         \hline 
       $e_{V_\text{cell}}$: model in \cite{Blondel_TCST_Journal_2019} [mV] & {9.35} & 13.92 & 11.02 & 15.16 \\
    $e_{V_\text{cell}}$: model (\ref{new_overallss}) [mV]& \textbf{7.00}  & \textbf{10.31} & \textbf{8.21} & \textbf{11.23} \\
    Improvement [\%] & \textbf{25.1} & \textbf{25.9} & \textbf{25.5} & \textbf{25.92} 

    \end{tabular}
    \vspace{0.4cm}
        \caption{MAE and RMSE of the output voltage given by the model in \cite{Blondel_TCST_Journal_2019} and model (\ref{new_overallss}).}
     \label{experrortable0}
      \vspace{-0.6cm}
    \end{table}
Next, we consider the observers designed in Section $\text{\rom{6}.B}$ of gain $L$ with the same initialization. However, in this section, we feed them with the measured input current and output voltage obtained by experimentation. To compare the output voltage and SOC estimated by both observers, we take as a reference the measured output voltage and the experimental SOC calculated by integration of the corrected measured current as follows 
\begin{equation}
    SOC_\text{exp}(t):=-\frac{1}{3600Q_\text{cell}}\int_0^t I_\text{cell, cor}(\tau)d\tau,
    \end{equation}
where $I_\text{cell, cor}$ is the corrected measured input current.

\begin{figure}[thpb]
    \centering
    \includegraphics[scale=0.24]{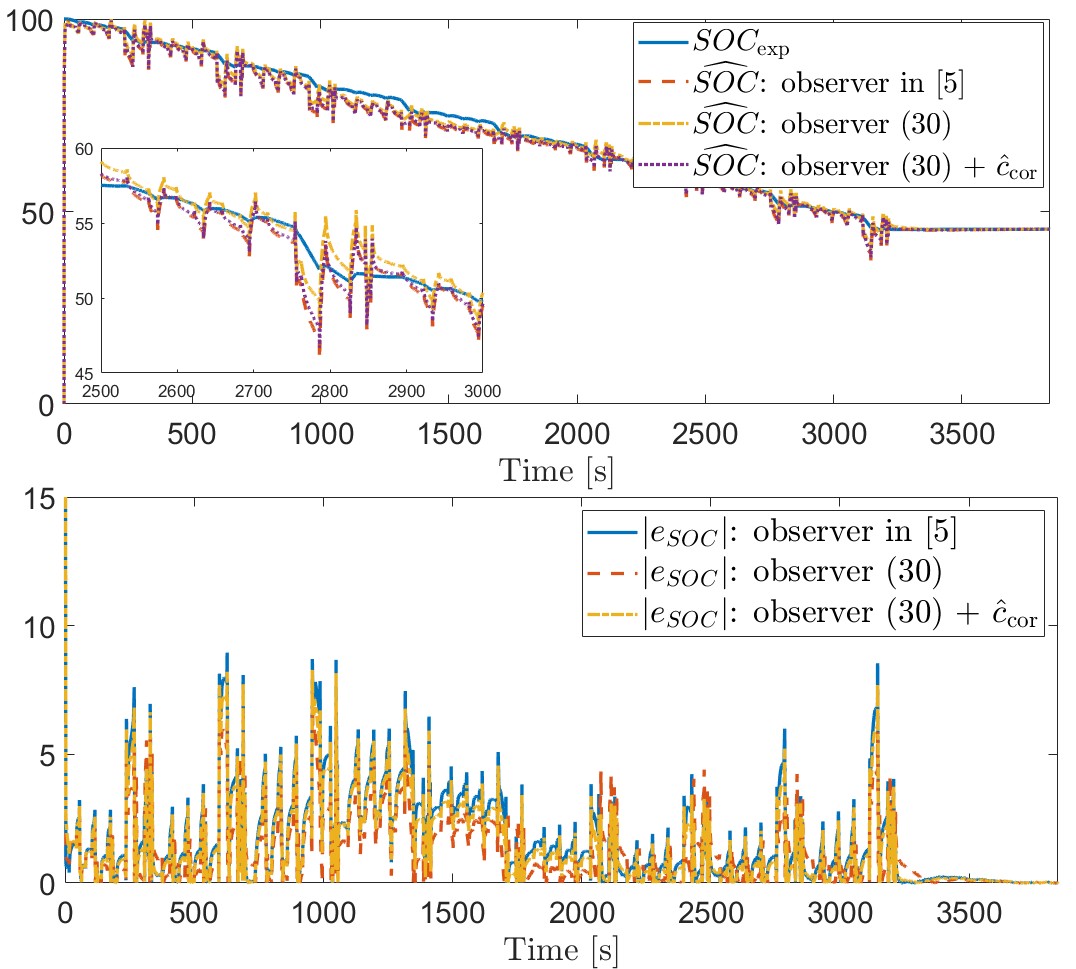}
    \vspace{-0.4cm}
    \caption{$SOC_\text{exp}$ and $\widehat{SOC}$ generated by the observer in \cite{Blondel_IFAC_Journal_2017}, observer (\ref{newobs}) and observer (\ref{newobs}) with $\hat{c}_\text{cor}$
(top) and norm of the estimation errors $e_{SOC} := SOC_\text{exp}-\widehat{SOC}$ (bottom).}
    \label{expsoc}
\end{figure}
Figure \ref{expsoc} shows the experimental SOC and the estimated ones, as well as the corresponding norm of the estimation error $e_{SOC}:=SOC_\text{exp}-\widehat {SOC}$. It is not clear which observer has the most accurate SOC. For this reason, we quantified the results by computing the average MAE and the average RMSE of the estimation errors for each of the observer in \cite{Blondel_IFAC_Journal_2017}, observer (\ref{newobs}) and observer (\ref{newobs}) with $\hat c_\text{cor}$ over the same 20 scenarios. The results are shown in Table \ref{experrortable}. The values in parenthesis represent the improvement of the corresponding observer with respect to the observer in \cite{Blondel_IFAC_Journal_2017}. Observer (\ref{newobs}) has the least SOC estimation error. Hence, observer (\ref{newobs}) based on the corrected model estimates the state of charge of lithium-ion batteries more accurately than the other observers. 

In addition, we also computed the MAE and RMSE of the voltage error between the measured output voltage and the estimated output voltages generated by the observer in \cite{Blondel_IFAC_Journal_2017} and observer (\ref{newobs}) averaged over the same 20 scenarios. The results are also shown in Table \ref{experrortable}. The voltage error associated to observer (\ref{newobs}) is lower than that of the model in  \cite{Blondel_IFAC_Journal_2017}. Hence, observer (\ref{newobs}) also estimates the output voltage more precisely than the observer in \cite{Blondel_IFAC_Journal_2017}.

\begin{table}[h]
    \centering
    \begin{tabular}{l|lll}
         & MAE & RMSE  &   \\ 
         \hline
       $e_{SOC}$: observer in \cite{Blondel_IFAC_Journal_2017} [\%]&  1.57 & 2.33  \\
      $e_{SOC}$: observer (\ref{newobs}) [\%]& \textbf{1.16 (26.1)} & \textbf{1.75 (24.9)} \\
      $e_{SOC}$: observer (\ref{newobs}) + $\hat{c}_\text{cor}$ [\%]& {1.40} & {2.11} \\
\hline
       $e_{{V}_\text{cell}}$: observer in \cite{Blondel_IFAC_Journal_2017} [mV]& 0.53 & 3.27\\
      $e_{{V}_\text{cell}}$: observer (\ref{newobs}) [mV]& \textbf{0.50 (5.7)} & \textbf{3.23 (1.2)} \\

   \end{tabular}
   \vspace{0.4cm}
    \caption{Average MAE and RMSE over 20 simulations of the SOC estimation errors and of the estimated concentrations error. The values in parenthesis represent the percentage of improvement with respect to the observer in \cite{Blondel_IFAC_Journal_2017}.}
    \label{experrortable}\vspace{-0.6cm}
\end{table}

\section{Conclusion}
We have presented an approach to correct the lithium concentrations of a finite-dimensional SPM model to asymptotically eliminate the errors induced by the PDE discretization for constant currents. As a result, more accurate variables are generated by the finite-dimensional model as illustrated in simulations. We have then exploited these corrections to derive a new output voltage equation and thus a new state space model. Two observer design strategies have been proposed for this new model, with robust stability guarantees. The estimated variables generated by the chosen observer are then corrected to also asymptotically match those of the PDEs. The obtained simulation and experimentation results show significant improvement in terms of state estimation.
Among the possible future works we can envision is the design of sampled-data observers for real-time implementation.


{\appendix[Proof of Lemma \ref{L3}]
We first show that all the eigenvalues of matrix $A_s$ are non-positive (note that these are real as $A_s$ is a tridiagonal matrix with symmetric coefficients signs). We invoke for this purpose Gersgorin disk theorem. This theorem states that each eigenvalue $\lambda_m$, with $m \in \{1,\hdots,N_s\}$, of $A_s$ satisfies at least one the inequalities $|\lambda_m-(A_s)_{ii}|\leq \rho_i$ with $\rho_i:=\sum_{\substack{j=1 \\ j\neq i }}^{N_s} |(A_s)_{ij}|$ for $i \in \{1,\hdots,N_s\}$. In view of the expression of $A_s$, this means that each $\lambda_m$ of $A_s$ satisfies at least one of the inequalities $|\lambda_m+\mu_1^s|\leq \mu_1^s$,  $|\lambda_m+\upsilon_i^s|\leq \upsilon_i^s$ for $i \in \{2,\hdots,N_s-1\}$ and $|\lambda_m+\Tilde{\mu}_{N_s}^s|\leq \Tilde{\mu}_{N_s}^s$. Hence, we deduce that $\lambda_m \leq 0$ for all $m \in \{1,\hdots,N_s\}$. On the other hand, by solving $A_sx=\textbf{0}_{N_s \times 1}$ for any $x \in \mathbb{R}^{N_s}$, we derive $\text{ker}(A_s)=\{\alpha \textbf{1}_{N_s \times 1}:\alpha \in \mathbb{R}\}$ is of dimension 1. Hence, the rank of $A_s$ is $N_s-1$, which means that there is a single eigenvalue of $A_s$ that is equal to 0. Consequently, $A_s$ admits $N_s-1$ strictly negative eigenvalues.

 Let $\lambda_m\in\mathbb{R}_{<0}$ be an eigenvalue of $A_s$, let $x_m$ be a corresponding non-zero eigenvector, i.e., $A_sx_m=\lambda_m x_m$ and $x_m\neq 0$. Given that $\Gamma_s A_s=\textbf{0}_{1 \times N_s}$, we derive $\Gamma_sA_sx_m=\Gamma_s\lambda_m x_m=\textbf{0}_{1 \times N_s}$. Thus, as $\lambda_m \neq 0$, we obtain 
\begin{equation}
     \sum_{i=1}^{N_s} V^s_ix_{m,i}=0.
     \label{51}
\end{equation}
On the other hand, $A_sx_m=\lambda_m x_m$ is equivalent to, for any $i \in \{1, \hdots, N_s\}$
\begin{equation}
    \sum_{j=1}^{N_s}(A_s)_{ij}x_{m,j}=\lambda_m x_{m,i}.
    \label{52}
\end{equation}
Using (\ref{51}), we obtain $x_{m,{N_s}}=-\frac{1}{V^s_{N_s}}\sum_{i=1}^{N_s-1} V^s_ix_{m,i}$. Hence, (\ref{52}) is equivalent to $\sum_{j=1}^{N_s-1}(A_s)_{ij}x_{m,j}+(A_s)_{iN_s}x_{m,{N_s}}=\sum_{j=1}^{N_s-1}(A_s)_{ij}x_{m,j}-\frac{1}{V^s_{N_s}}\sum_{j=1}^{N_s-1} (A_s)_{iN_s}V^s_jx_{m,j}=\lambda_m x_{m,i}$.
Therefore, in view of the definition of $\tilde A_s$ we derive for any $i \in \{1, \hdots, N_s-1\}$
\begin{equation}
    \sum_{j=1}^{N_s-1}((A_s)_{ij}-(A_s)_{iN_s}\frac{V^s_j}{V^s_{N_s}})x_{m,j}=\sum_{j=1}^{N_s-1}(\tilde A_s)_{ij}x_{m,j}=\lambda_m x_{m,i}.
    \label{53}
\end{equation}
From (\ref{53}), we obtain $\tilde A_s \tilde x_m =\lambda_m  \tilde x_m $, where $\tilde x_m$ is the vector of the $N_s-1$ first coefficients of $x_m$. Therefore, $\lambda_m$ is an eigenvalue of $\tilde A_s$. Consequently, any of the $N_s-1$ strictly negative eigenvalue of $A_s$ is also an eigenvalue of $\tilde A_s$. Since $\tilde A_s$ is of dimension $N_s-1\times N_s-1$, this implies that all the eigenvalues of $\tilde A_s$ are strictly negative: matrix $\tilde A_s$ is Hurwitz. Lemma \ref{L3} is thus proven.


\bibliographystyle{plain}
\bibliography{IEEEabrv,Journal.bib}


 
\vspace{11pt}



\vfill

\end{document}